\documentclass[preprint,12pt,sort&compress]{elsarticle}
\usepackage[left=2.5cm,right=2.5cm, top=2.3cm, bottom=2.3cm]{geometry}

\usepackage{amssymb,amsfonts}
\usepackage{amsmath,bm,mathrsfs,amsthm}
\usepackage{algorithm}
\usepackage{multirow}
\usepackage{enumitem}
\usepackage{booktabs}
\usepackage[title]{appendix}%
\newtheorem{res}{Result}
\usepackage{float}
\usepackage{epsfig}
\usepackage[caption=false]{subfig}

\journal{}

\begin{document}

\begin{frontmatter}



\title{Exponential-polynomial divergence based inference for nondestructive one-shot devices under progressive stress model} 


\author[inst1]{Shanya Baghel}

\affiliation[inst1]{organization={Department of Mathematics and Computing, Indian Institute of Technology (ISM) Dhanbad},
            postcode={826004}, 
            state={Jharkhand},
            country={India}}

\author{Shuvashree Mondal\corref{cor1}\fnref{inst1}}
\ead{shuvasri29@iitism.ac.in}
 \cortext[cor1]{Shuvashree Mondal}

\begin{abstract}
Nondestructive one-shot device (NOSD) testing plays a crucial role in engineering, particularly in the reliability assessment of high-stakes systems such as aerospace components, medical devices, and semiconductor technologies.  Accurate reliability prognosis of NOSD testing data is essential for ensuring product durability, safety, and performance optimization.  The conventional estimation methods like maximum likelihood estimation (MLE) are sensitive to data contamination, leading to biased results.  Consequently, this study develops robust inferential analysis for NOSD testing data under a progressive stress model.  The lifetime of NOSD is assumed to follow Log-logistic distribution.  The estimation procedure addresses robustness by incorporating Exponential-polynomial divergence (EPD). 
 Equipped with three tuning parameters, EPD based estimation is proven to be more flexible than density power divergence estimation frequently used for one-shot device testing data analysis.  Further, we explore the asymptotic behaviour of minimum EPD estimator (MEPDE) for large sample size.  The robustness of MEPDE is analytically studied through influence function.  Since tradeoff between efficiency and robustness of EPD based estimation is governed by three tuning parameters, a novel approach leveraging Concrete Score Matching (CSM) is introduced to optimize the tuning parameters of MEPDE.  Moreover, a comparative study with the existing methods of finding tuning parameters is conducted through extensive simulation experiment and data analysis.  Another aspect of this study is determining an optimal plan to ensure a successful ALT experiment within specified budget and time constraints.  It is designed on A-optimality criteria subject to the given constraints and is executed using the constraint particle swarm optimization (CPSO) algorithm.
\end{abstract}



\begin{keyword}
Nondestructive one-shot device\sep Progressive stress \sep Exponential-polynomial divergence \sep Robust estimation \sep Concrete score matching \sep Constraint particle swarm optimization. 

\end{keyword}

\end{frontmatter}



\section{Introduction}
Nondestructive one-shot device (NOSD) testing is a critical tool in engineering, particularly in the reliability prognosis of high-stakes systems such as aerospace components \cite{luan2023life}, medical devices \cite{shrivastava2014detection}, semiconductor technologies \cite{nguyen2020enhance} and automotive safety systems \cite{abdullah2025adaptive}.  NOSD can withstand multiple tests, offering additional data for reliability analysis, contrary to one-shot devices that are immediately destroyed after testing.  Units such as brake or air conditioning fluids in cars, fan condensers, metal fatigue, light bulbs, circuit breakers, pressure relief valves and stability of pharmaceuticals are examples of NOSD \cite{balakrishnan2023robust}.  For NOSD, often observations are limited to recording whether a failure occurs before or after a specified inspection time, resulting in the analysis of dichotomous data only.  Furthermore, highly reliable products have become the standard due to the competitive market environment.  To perform reliability analysis of such products within a specified time and budget, accelerated life testing (ALT) experiments are conducted; readers may refer to \cite{sharma2018hierarchical,sharma2021hierarchical,prajapati2023misspecification,ashkamini2023bayes,liao2023accelerated,zhang2024reliability,kateri2024product,nelson2024advances,qin2024reliability,kumari2024bayes,salah2025point} and references therein.  ALT can be classified according to the stress incorporation as constant stress, step-stress, progressive stress and cyclic stress, one may see \cite{nelson2009accelerated} for more information on it.  Zhuang et al. \cite{zhuang2023data} stated that  among these methods, progressive stress ALT (PSALT) is the most effective and flexible in reducing the life span of any product.  

In PSALT, the stress on test units increases continuously over time. 
 For example, in bridge load testing, engineers progressively increase the weight or load on a bridge to determine its maximum capacity and identify potential failure points \cite{gangone2014development,tonelli2023prestressed,savino2023large}.  In aircraft wing stress testing, the wings of an aircraft are subject to progressive stress to simulate extreme conditions and ensure they can withstand real-world forces \cite{godines2017damage,dormohammdi2017damage}.  PSALT with linearly increasing stress is referred to as ramp-stress ALT.  For more information on PSALT, readers may refer to \cite{rong2004statistical,abdel2007progressive,abdel2011inference,ismail2022progressive,mahto2024bayesian} and references therein.  Existing studies on the reliability analysis of NOSD are concentrated towards step-stress ALT (SSALT) experiments as evidenced by works of \cite{balakrishnan2022robustr,balakrishnan2022restricted,balakrishnan2023robust,balakrishnan2024non} and references cited therein. 
 However, the reliability prognosis of NOSD under PSALT remains unexplored.

The lifetime of NOSD testing units are commonly modeled by exponential, Weibull or lognormal distribution in the literature.  There exist numerous real-world occurrences in which the lifetime of units follows a distribution different from these widely used ones.  For instance, Log-logistic lifetime distribution is also used in survival analysis to model ALT data \cite{klein2006survival}.  One can refer to \cite{mahto2024bayesian} and references therein to go through the utility of Log-logistic distribution in modelling lifetime data.  Its probabilistic behaviour resembles generalized Pareto distribution and Burr distribution \cite{tadikamalla1980look}.  In this study lifetime of NOSD is assumed to follow Log-logistic distribution since it is pretty versatile as its hazard rate can be decreasing or inverted bathtub-shaped \cite{johnson1995continuous}.

When the failure mechanism of the data aligns with the assumed model assumptions, conventional maximum likelihood estimation (MLE) performs efficiently. However, its performance deteriorates significantly in the presence of contaminated data, where the failure mechanisms of some NOSD testing units slightly deviate from the assumed model. Since purely uncontaminated data is rare in practice, it is crucial to develop robust estimation methods that remain stable and accurate even when the data contains contamination. Such methods are essential for ensuring robust reliability prognosis in real-world scenarios, where deviations from model assumptions are inevitable.  It is deciphered from the literature that density power divergence (DPD) \cite{basu1998robust} based estimation has been frequently exploited for robust inferential analysis; readers may refer to \cite{bp(2019),balakrishnan2019robust1,balakrishnan2019robust2,blcamapa2020,castilla2021robust,balakrishnan2023robust1,balakrishnan2024robust,balakrishnan2024step} and references cited therein.  The tuning parameter $\gamma$ in DPD based estimation (DPDE) coordinates robustness and efficiency.  A higher value of $\gamma$ enhances robustness by reducing sensitivity to outliers, while a lower value improves efficiency under standard model assumptions.  Mukherjee et al. \cite{mukherjee2019b} introduced the Bregman-Exponential Divergence (BED) with the tuning parameter $\alpha$ wherein the corresponding estimating equation can be interpreted as a weighted likelihood estimating equation.  Furthermore, Singh et al. \cite{singh2021robust} introduced a generalized divergence known as Exponential-polynomial divergence (EPD) that encompasses both DPD and BED as special cases.  The tuning parameters $(\alpha,\beta,\gamma)$ in EPD enable greater flexibility in adjusting robustness and efficiency than DPD and BED.  Therefore, this study proposes incorporating EPD for a robust estimation procedure for NOSD testing units.  Additionally, the asymptotic distribution of minimum EPD estimate is thoroughly developed based on the discussion of Calvino et al. \cite{calvino2021robustness}.  In conjunction with this, the influence function analysis is carried out to assess the robustness of the MEPDE analytically, providing deeper insights into its stability and performance.

The tuning parameters $(\alpha,\beta,\gamma)$ in EPD based estimation are crucial in balancing robustness and efficiency.  These parameters vary across a continuous range, allowing countless alternatives in the estimation procedure.  However, in any real-world scenario, the experimenter must select most suitable value of these tuning parameters for a specific dataset.  Thus, finding optimal tuning parameters is very essential for such studies.  In this direction, method proposed by Warwick and Jones (WJ) \cite{warwick2005choosing} and its improved version iterative Warwick and Jones (IWJ) algorithm \cite{basak2021optimal} are frequently used.  Castilla and Chocano \cite{castilla2022choice} suggested three methods for determinig tuning parameter which are at least as efficient as IWJ against high contamination while avoiding excessive computational overhead.  Another significant contribution is the study of Sugasawa and Yonekura \cite{sugasawa2021selection,yonekura2023adaptation} who focused on robustness and propsed a method based on Hyvarinen score matching (HSM) \cite{dawid2015bayesian,shao2019bayesian}. HSM is a simple technique for estimating non-normalized models \cite{hyvarinen2005estimation}.  Extending HSM to the discrete data setting is not straightforward, though. 
 Moreover due to the dichotomous nature of NOSD data, implementing HSM based optimization for tuning parameters \cite{sugasawa2021selection} has proven to be challenging. 
 To address this, a generalized score matching approach, such as concrete score matching (CSM) \cite{meng2022concrete}, which is specifically designed for the discrete data setting, can be utilized.  The idea is to construct a neigbourhood structure concerning the data as a substitute for the gradient through local directional changes.  This study proposes a novel approach in finding optimal tuning parameters by leveraging CSM \cite{meng2022concrete}.

 Another signifcant aspect of this study is to develop an optimal plan.  The aim is to find a suitable sample allocation and inspection times such that the trace of the asymptotic covariance matrix of estimates is minimized subject to given constraints.  Meta-heuristic Particle swarm optimization (PSO) \cite{kennedy1995particle} is used in developing an optimal design where constraints are handled using Deb's rule \cite{deb2000efficient}.  This approach of constraint particle swarm optimization (CPSO) was considered by Ang et al. \cite{ang2020constrained}.  While most studies in the literature have concentrated on optimal plans for destructive one-shot devices \cite{balakrishnan2014best,ling2019optimal,ling2020optimal,lee2020optimal,balakrishnan2022optimal,ling2022optimal,ling2024optimal}, this study proposes an optimal plan for NOSD testing units.  

This study presents a robust inference of NOSD under the PSALT model, assuming that the lifetime of NOSD follows a Log-logistic distribution. The estimation procedure is based on Exponential Polynomial Divergence (EPD) \cite{singh2021robust}.  The asymptotic distribution of the minimum EPD estimate (MEPDE) is derived following the approach of Calvino et al. \cite{cal2021}.  Robustness is examined analytically through influence function analysis.  Additionally, the study proposes utilizing Concrete Score Matching (CSM) \cite{meng2022concrete} to determine optimal tuning parameters. Finally, an optimal design strategy is developed to identify the best sample allocation and inspection times, ensuring that the experiment is completed within the given budget and time constraints.

 The rest of the article proceeds as follows.  Section \ref{sec2} focuses on building a progressive stress model.  The Exponential-polynomial divergence based estimation is discussed in Section \ref{sec3}.  Section \ref{sec4} discusses methods of obtaining the optimal value of tuning parameters.  Section \ref{sec5} takes on numerical study based on previous sections.  Finally, Section \ref{sec6} contains the optimal design of the PSALT experiment.  Concluding remarks are given in Section \ref{sec7}.

\section{Progressive stress model}\label{sec2}
\noindent A model setup is required to execute a successful reliability prognosis for nondestructive one-shot device (NOSD) testing data. This study is carried out within the framework of progressive stress accelerated life testing (PSALT) model. The PSALT is developed under the following assumptions 
\subsection{Assumptions}
\begin{itemize}
    \item The lifetime of NOSD follows the Log-logistic distribution with shape parameter $\mu$ and scale parameter $\lambda$ whose cdf and pdf are given as
    \begin{align*}
        F(t;\mu,\lambda)&=\left[1+\left(t/\lambda\right)^{-\mu}\right]^{-1}.\\
        f(t;\mu,\lambda)&=\frac{(\mu/\lambda)(t/\lambda)^{\mu-1}}{\left[1+(t/\lambda)^{\mu}\right]^2}\;,\;t>0\,(\mu,\lambda)>0.
    \end{align*}
    \item The scale parameter $\lambda$ satisfies the inverse power law as
    \begin{equation*}
        \lambda(t)=\frac{1}{a[s(t)]^b}\;,\;(a,b)>0.
    \end{equation*}
    \item The progressive stress s(t) is directly proportional to time, i.e.
    \begin{equation*}
        s(t)=\nu t\;,\;\nu>0,
    \end{equation*}
    where $\nu$ is the stress rate.
    \item Tampered failure rate model (TFRM) is considered for
modelling the effect of stress change \cite{rong2004statistical}.
\end{itemize}

\begin{figure}[htb!]
\begin{center}
\subfloat[Progressive stress\label{st}]{\includegraphics[height=4.5cm,width =0.43\textwidth]{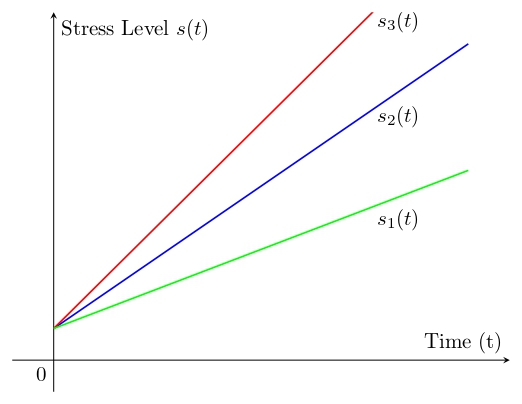}} 
\subfloat[Hazard rate\label{haz}]{\includegraphics[height=4.6cm,width =0.43\textwidth]{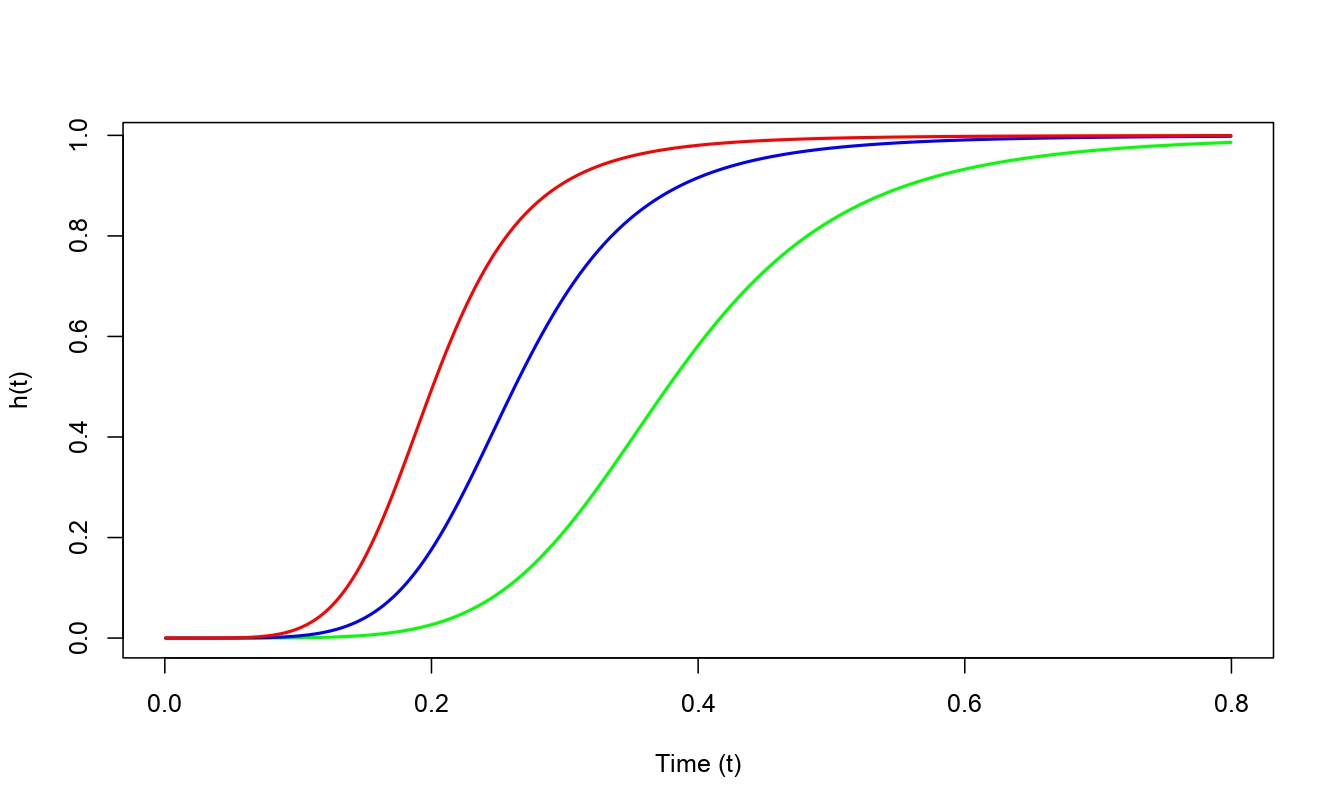}}
\end{center}
\caption{Progressive stress (left) and Hazard rate (right) for Log-logistic lifetime distribution under TFRM with three stress levels.}
\label{hazst}
\end{figure}

For the testing procedure, the $N,$ NOSD units are divided into $k(\geq 2)$ groups under an accelerated life testing experiment. Each group contains $N_i$ units and is subject to progressively increasing stress. For $i=1,2,\dots,k\;;$ $s_i(t)=\nu_i t.$ Therefore, the hazard rate and cumulative hazard rate are given below.
\begin{align*}
   h_i(t)&=\frac{\mu(a\nu_i^b)^{\mu}t^{\mu(b+1)}}{1+(a\nu_i^b)^{\mu}t^{\mu(b+1)}}.\\
   H_i(t)&=\frac{1}{b+1}ln\,\left[1+(a\nu_i^b)^{\mu}t^{\mu(b+1)}\right]\;;\;i=1,2,\dots,k..
\end{align*}
The survival function under TFRM incorporating progressive stress can be expressed as
\begin{equation}
    S_i(t)=\left[1+(a\nu_i^b)^{\mu}t^{\mu(b+1)}\right]^{-1/b+1}\;;\;i=1,2,\dots,k.
\end{equation}
Here, the model parameters to be estimated can be denoted as $\bm{\theta}=(a,b,\mu).$  Figure \eqref{hazst}  depicts a progressive stress setup and hazard rate for Log-logistic lifetime distribution under TFRM with three stress levels.

The NOSD units are inspected at the ordered inspection time points $\tau_{i1}\leq \tau_{i2}\leq\dots\tau_{ij}\leq\dots\tau_{iJ_i}$ for the $i$th group where $i=1,2,\dots,k$ and $\tau_{i0}=0$.  The experiment for each group is terminated at $\tau_{iJ_i}$. 
 Let $t_i$ denotes lifetime of any NOSD unit in $ith$ group. The failure probability $p_{ij}(\bm{\theta}){=}p_{ij}$ in the interval $(\tau_{i(j-1)},\tau_{ij})$ and survival probability $p_{is}(\bm{\theta}){=}p_{is}$ are given as,
\begin{align}
    p_{ij}&{=}P(\tau_{i(j-1)}< T\leq\tau_{ij}){=}\left[1+(a\nu_i^b)^{\mu}\tau_{i(j-1)}^{\mu(b+1)}\right]^{-\frac{1}{b+1}}\notag\\
    &\quad{-}\left[1+(a\nu_i^b)^{\mu}\tau_{ij}^{\mu(b+1)}\right]^{-\frac{1}{b+1}},\label{fprob}\\
    p_{is}&=S_i(\tau_{iJ_i})=\left[1+(a\nu_i^b)^{\mu}\tau_{iJ_i}^{\mu(b+1)}\right]^{-\frac{1}{b+1}}.\label{sprob}
\end{align}
If $n_{ij}$ is the number of observed failures at the $j^{th}$ inspection time point of the $i^{th}$ stress rate group, the number of failures at the $i^{th}$ group is given by $n_i=\sum_{j=1}^{J_i}n_{ij}.$
Thus, the likelihood function and maximum likelihood estimate (MLE) are given as follows.
\begin{align}
    L(\bm{\theta})&\propto \prod_{i=1}^{k}\prod_{j=1}^{J_i}p_{ij}^{n_{ij}}p_{is}^{N_i-n_i},\\
    \hat{\bm{\theta}}_{\text{\tiny MLE}}&=arg\mathop{max}_{\bm{\theta}}ln\,L(\bm{\theta}).
\end{align}
MLE works efficiently but lacks robustness.  Therefore, we have adopted a generalized divergence measure named Exponential-Polynomial divergence (EPD) \cite{singh2021robust} for the estimation procedure \cite{kim2024robust}.

\section{Exponential-polynomial divergence}\label{sec3}
Exponential-polynomial divergence (EPD) was proposed by Singh et al.\cite{singh2021robust} as a generalized class of Bregman divergence.  Under the assumptions stated by Bregman \cite{bregman1967relaxation}, the Bregman divergence for two density functions $g$ and $f$ takes the form
\begin{align*}
   D_B(g; f) &= \int_x \bigg[B\big(g(x)\big) - B\big(f(x)\big) \bigg. \bigg.- \Big\{g(x) - f(x)\Big\} B'\big(f(x)\big)\bigg]\,dx.
\end{align*}
Here, the function $B$ is strictly convex and $B^{'}$ is the derivative of $B$ concerning its argument.  Because of the linear characteristic of the integral, $B$ is not uniquely determined.  Several choices of convex function $B$ result in different special cases of Bregman divergence; readers may refer to \cite{basak2022extended,basak2024extended} and references therein.  Singh et al. \cite{singh2021robust} leveraged this non-uniqueness by proposing a convex function $
    B(x)=\frac{\beta}{\alpha^2}\big(e^{\alpha x}-1-\alpha x\big)+\frac{1-\beta}{\gamma}\big(x^{\gamma+1}-x\big)\,;\alpha\in \mathcal{R}, \beta\in [0,1], \gamma\geq 0
$, and introduced a divergence measure called the Exponential-polynomial divergence (EPD).  Here, $\alpha$, $\beta$, $\gamma$ are tuning parameters.  The class of divergence measures for particular values of $(\alpha,\beta,\gamma)$ is given in the table \ref{tuneidea}.

\begin{table}[htb!]
\caption{Class of divergence measures.}
\begin{center}
{\scalebox{0.8}{
\begin{tabular}{ll}
\hline%
$(\alpha,\beta,\gamma)$&\textbf{Divergence measure}\\
\hline
$\beta=0$& Density Power Divergence (DPD) \cite{basu1998robust} with parameter $\gamma$\\
$\beta=1$ & Brègman exponential-divergence (BED) \cite{mukherjee2019b} with parameter $\alpha$\\
$\beta\neq 0$& Combination of BED and DPD\\
$\beta=0$, $\gamma\to 0$& Kullback-Leibler (KL) divergence \cite{kullback1951information}\\
\hline
\end{tabular}}}
\end{center}
\label{tuneidea}
\end{table}

In the context of NOSD testing, EPD divergence is defined between theoretical and empirical failure-survival probabilities. 
 Define empirical failure and survival probabilities as $\frac{n_{ij}}{N_i}{=}q_{ij}$ and $\frac{N_i-n_i}{N_i}{=}q_{i(J_i+1)}$ respectively.  The theoretical failure and survival probabilities are given in the equations \eqref{fprob} and \eqref{sprob}, where we denote $p_{is}=p_{i(J_i+1)}.$ 
 Then, EPD for NOSD testing units is explained by the expression
\begin{align}
    D_{EP}(\bm{\theta})&=\sum_{i=1}^{k}\sum_{j=1}^{J_i+1}\Bigg[(1-\beta)p_{ij}^{\gamma+1}+\frac{\beta}{\alpha}e^{\alpha p_{ij}}\left(p_{ij}-\frac{1}{\alpha}\right)\Bigg. -\left\{\frac{\beta}{\alpha}e^{\alpha p_{ij}}+\frac{\gamma+1}{\gamma}\right.\bigg.\times(1-\beta)p_{ij}^{\gamma}\bigg\}q_{ij}\notag\\
    &\qquad\qquad\qquad +\left.\left\{\frac{\beta}{\alpha^2}e^{\alpha q_{ij}}+\frac{1-\beta}{\gamma}q_{ij}^{\gamma+1}\right\}\right].\label{epd}
\end{align}
The EPD ignoring the terms independent of parameters is given as,
\begin{align}
    D_{EP}(\bm{\theta})&=\sum_{i=1}^{k}\sum_{j=1}^{J_i+1}\Bigg[(1-\beta)p_{ij}^{\gamma+1}+\frac{\beta}{\alpha}e^{\alpha p_{ij}}\left(p_{ij}-\frac{1}{\alpha}\right)\Bigg.-\left\{\frac{\beta}{\alpha}e^{\alpha p_{ij}}+\frac{\gamma+1}{\gamma}\times(1-\beta)p_{ij}^{\gamma}\bigg\}q_{ij}\right].\label{epd-}
\end{align}
Therefore, the minimum EPD estimate (MEPDE) can be obtained as
\begin{equation}
\hat{\bm{\theta}}_{\scriptscriptstyle EP}=arg\mathop{min}_{\bm{\theta}} D_{EP}(\bm{\theta}).
\end{equation}

\begin{res}\label{res1}
The set of estimating equations for obtaining MEPDE for NOSD testing data takes the following form.
\begin{equation}
\sum_{i=1}^{k}\sum_{j=1}^{J_i+1}\bigg[p_{ij}\Big\{(1-\beta)(\gamma+1)p_{ij}^{\gamma-1}+\beta e^{\alpha p_{ij}}\Big\}\big(q_{ij}-p_{ij}\big)\bigg]u_{ij}{=}0,\label{mepde}
    \end{equation}
where, $u_{ij}=\frac{\partial}{\partial\bm{\theta}}ln\,p_{ij}$, $\alpha\in \mathcal{R}$, $\beta\in [0,1]$ and $\gamma\geq 0$.   The expressions $u_{ij}$ for $j=1,2,\dots, J$ and $u_{i(J_i+1)}=u_{is}$ are described in the appendix. 
\end{res}
The tuning parameters $(\alpha,\beta,\gamma)$ play a crucial role in estimation as it balances efficiency and robustness.  Therefore, a discussion on optimal tuning parameters is separately discussed in Section \ref{sec4}.  Further, to examine the asymptotic behaviour of MEPDE, the following theorem motivated by the study of Calvino et al. \cite{cal2021} is presented.
\begin{res}\label{res2}
  Let $\bm{\theta}_0$ be true value of parameter 
$\bm{\theta}.$ The asymptotic distribution of the MEPDE ${\hat{\bm{\theta}} }_{\scriptscriptstyle EP}$ is given by
\begin{equation*}
  \sqrt{N}\Big({\hat{\bm{\theta}} }_{\scriptscriptstyle EP} - {\bm{\theta}_0}\Big) \xrightarrow[N\to \infty]{\mathscr{L}}Nor\Big(\bm{0}_{3}, J^{-1}_{\scriptscriptstyle EP}({\bm{\theta}_0}) K_{\scriptscriptstyle EP}({\bm{\theta}_0})J^{-1}_{\scriptscriptstyle EP}({\bm{\theta}_0}) \Big), 
\end{equation*}
\end{res}
\begin{proof}
The description of the notation and proof of the result is given in the appendix.
\end{proof}

\subsection{\textbf{Property of robustness}}
This subsection includes robustness analysis through influence function (IF).  Suppose, for a true distribution $\bm{G}$, functional of any estimator is denoted by $T_{EP}(\bm{G}).$  Then, the influence function is defined as
\begin{equation*}
    IF(t;T_{\scriptscriptstyle EP},\bm{G}){=}\lim_{\epsilon \to 0}\frac{T_{\scriptscriptstyle EP}(\bm{G}_{\varepsilon})-T_{\scriptscriptstyle EP}(\bm{G})}{\epsilon}{=}\left.\frac{\partial(T_{\scriptscriptstyle EP}(\bm{G}_{\epsilon}))}{\partial\epsilon}\right\vert_{\epsilon{\to}0^{+}}.
\end{equation*}
Here, $\bm{G}_{\epsilon}=(1-\epsilon)\bm{G}+\epsilon\delta_{\bm{t}}$ is the contaminated model where $\epsilon,$ $(0<\epsilon<1)$ is the proportion of contamination and $\delta_{\bm{t}}$ denotes the degenerate distribution at point $\bm{t}$.  
\begin{res}\label{res3}
  The Influence function of $\hat{\bm{\theta}}_{\scriptscriptstyle EP}$ for NOSD testing data is obtained as
  \begin{align}
    &IF(\bm{t_1},\dots,\bm{t_k};T_{\scriptscriptstyle EP},\bm{G})=J^{-1}_{\scriptscriptstyle EP}(\bm{\theta})\sum_{i=1}^{k}\sum_{j=1}^{J_i+1}\left[p_{ij}\Big\{(1-\beta)\Big.\right.\left.\Big.(\gamma+1)p_{ij}^{\gamma-1}+\beta e^{\alpha p_{ij}}\Big\}\right.\Big.\left(\delta_{\bm{t_i}}-p_{ij}\right)\Big]u_{ij},
  \end{align}
  where $\delta_{\bm{t_i}}$ is the degenerate distribution at the outlier point $\bm{t_i}=(t_{i1},\dots,t_{iJ_i+1})^T\in\{0,1\}^{J_i+1}$ with $\sum_{j=1}^{J_i+1}t_{ij}=1.$
  \end{res}
\begin{proof}
The proof of result and description of expressions are given in the appendix.
\end{proof}
The IF for MEPDE is bounded for any finite $\alpha,$ $\gamma>0,$ and $\beta\in[0,1]$.  A smaller IF value indicates a more robust estimator.

\section{Optimal tuning parameters}\label{sec4}
It is evident from Section \ref{sec3}, that tuning parameters $(\alpha,\beta,\gamma)$ play a crucial role in balancing robustness and efficiency in the estimation procedure based on the EPD measure.  These parameters vary across continuous range $\alpha\in \mathcal{R}$, $\beta\in [0,1]$ and $\gamma\geq 0$ allowing countless alternatives in estimation procedure.  However, in any real-world scenario, the experimenter must select most suitable value for the tuning parameter based on the specific dataset.  Singh et al.\cite{singh2021robust} have used Warwick and Jones method (WJ)\cite{warwick2005choosing} following the approach of Ghosh and Basu \cite{gbas2015} for the tuning in EPD estimation.  In WJ, the objective is to minimize the asymptotic mean-square-error (MSE) of MEPDE $\hat{\bm{\theta}}_{EP},$ which is given by
\begin{align*}
&\widehat{MSE}(\alpha,\beta,\gamma)=(\hat{\bm{\theta}}_{EP}-\bm{\theta}_P)^T(\hat{\bm{\theta}}_{EP}-\bm{\theta}_P)+\frac{1}{N} Tr\Big(J_{EP}({\hat{\boldsymbol{\theta}}_{EP}})^{-1} K_{\beta}({\hat{\boldsymbol{\theta}}_{EP}})J_{\beta}({\hat{\boldsymbol{\theta}}_{EP}})^{-1}\Big),   
\end{align*}
where $\bm{\theta}_P$ is the pilot estimator and $Tr()$ is the trace of a variance-covariance matrix. The major drawback is that the outcome heavily relies on a pilot estimator.  An improvement of this method was suggested by Basak et al.\cite{basak2021optimal}, where they proposed an iterative procedure to eliminate the dependency on pilot estimator.  In this iterative Warwick and Jones algorithm (IWJ), the pilot estimator is replaced at each iteration with the updated estimator computed using the optimal values of $(\alpha,\beta,\gamma)$ at the particular iteration.  This process continues until stabilization is achieved.  We numerically observed that since MEPDE involved three tuning parameters, continuing the algorithm for the iterative process till stabilization was extremely time-consuming, bringing a disadvantage to this method for our model.

Castilla and Chocano suggested other crucial work in this direction \cite{castilla2022choice}.  Instead of minimizing the expected MSE, they focused on minimizing an estimated error between empirical probabilities $\frac{n_{ij}}{Ni}$ and theoretical probabilities $p_{ij}$.  Instead of a pilot estimator, grids of tuning parameters $(\alpha,\beta,\gamma)$ are used.  They observed that the suggested methods were at least as efficient as IWJ against high contamination while avoiding excessive computational overhead.  They proposed the following three methods.
\begin{itemize}
    \item Minimize the maximum of the absolute errors, $$\text{minAMAX} = \min\limits_{(\alpha, \beta, \gamma)} \max_{\substack{i=1,\dots,k \\ j=1,\dots,J_i+1}} \Bigg[\left\vert p_{ij} - \frac{n_{ij}}{N_i} \right\vert\Bigg].$$
    \item Minimize the mean absolute error,
    $$
    \text{minMAE}=\min\limits_{(\alpha, \beta, \gamma)}\frac{\sum_{i=1}^{k}\sum_{j=1}^{J_i+1}\left\vert p_{ij}-\frac{n_{ij}}{Ni}\right\vert}{\sum_{i=1}^{k}J_i+1}.
    $$
    \item Minimize the median of the absolute errors,
    $$
   \text{minAMED} = \min\limits_{(\alpha, \beta, \gamma)} \mathop{\text{median}}_{\substack{i=1,\dots,k \\ j=1,\dots,J_i+1}} \Bigg[\left\vert p_{ij} - \frac{n_{ij}}{N_i} \right\vert\Bigg].
    $$
\end{itemize}
Sugasawa and Yonekura \cite{sugasawa2021selection,yonekura2023adaptation} focused on robustness and proposed selection criteria based on Hyvarinen score matching \cite{dawid2015bayesian,shao2019bayesian}.  Since the present study revolves around dichotomous data, this method could not be applied in this study.  Therefore, we suggest a tuning parameter optimization method based on concrete score matching (CSM) \cite{meng2022concrete} specifically generalized for discrete data.

\subsection{\textbf{Optimal tuning parameter based on concrete 
score matching}}
Meng et al.\cite{meng2022concrete} developed a concrete score, a generalized score amenable to discrete and continuous data types.  The approach was aimed to construct a surrogate for the gradient by exploiting local directional changes to the inputs.  This can be done by leveraging connected neighbourhood structures tailored to the specific characteristics of the data.  Building on this foundation, the concrete score matching (CSM) method is developed here to find optimal value of tuning parameters $(\alpha, \beta, \gamma)$ for NOSD lifetime analysis.

For any NOSD in $i$th group failing within the interval $(\tau_{i(j-1)},\tau_{ij}],$ we generate a vector $X_{ij}=(x_{ij1},\dots,x_{ijj^{'}},\dots,x_{ij(J_i+1)})^{T}$ such that each element $x_{ijj^{'}}$ of vector $X_{ij}$ is explained as,
$$x_{ijj^{'}} = 
\begin{cases}
1 \;\text{if}\; j=j^{'}\\
0 \; \text{otherwise}\\
\end{cases}.
$$
Therefore, the sample space for $i$th group is defined as $\mathcal{X}_{i}{=}\{X_{ij}{:}j{=}1,{\dots},J_i{+}1\},$ for $i=1,\ldots k.$  Based on EPD, on each $X_{ij},$ we define the exponential of the maximizer of MEPDE as follows,
\begin{align}
Q_{\bm{\theta}}(X_{ij})&=exp\left[-\frac{1}{N_i}\sum_{l=1}^{N_i}\left(\sum_{j^{'}=1}^{J_i+1}\left\{(1-\beta)p_{ij^{'}}^{\gamma+1}+\frac{\beta}{\alpha}e^{\alpha p_{ij^{'}}}\right.\right.\right.\notag\\
&\qquad\Bigg.\left.\left(p_{ij^{'}}-\frac{1}{\alpha}\right)-\left(\frac{\beta}{\alpha}e^{\alpha p_{ij^{'}}}+\frac{\gamma+1}{\gamma}\times(1-\beta)p_{ij^{'}}^{\gamma}\bigg)x_{ijj^{'}}\right\}\Bigg)\right].\label{conq}
\end{align}
The approach is to deem $ Q_{\bm{\theta}}(X_{ij})$ as an unnormalized statistical model \cite{sugasawa2021selection,g2016}.  Jewson and Rosesell \cite{jewson2022general} also highlighted that the role of such unnormalized models can be explained in relative probability.  Following the assumptions outlined by Meng et al.\cite{meng2022concrete}, the weakly connected neighbourhood structure for NOSD testing data is constructed.  Let $\mathcal{N}$ be a function mapping each $X_{ij}$ to a set of neighbours such that 
\begin{flalign*}
 &\mathcal{N}(X_{i1})=\{X_{i2}\},\\
 & \mathcal{N}(X_{ij})=\{X_{i(j-1)},X_{i(j+1)}\}\,;\;j{=}2,\dots,J_i,\\
&  \mathcal{N}(X_{i(J_i+1)})=\{X_{i(J_i)}\}.
\end{flalign*}
Then concrete score on $X_{ij}$ is defined as,
\begin{align}
c_{data}(X_{ij},\mathcal{N})&\triangleq \Bigg[\frac{Q_{data}(X_{i(j-1)})-Q_{data}(X_{ij})}{Q_{data}(X_{ij})},\Bigg.\Bigg.\frac{Q_{data}(X_{i(j+1)})-Q_{data}(X_{ij})}{Q_{data}(X_{ij})}\Bigg]^{T},\label{cs}
\end{align}
for $i=1,2,\dots,k; j=1,2,\dots,J_i+1.$  Following the approach of Meng et al. \cite{meng2022concrete}, for any $ith$ group, we define the optimizing criteria given as, 

\begin{align}
&\Phi^{(i)}_{CSM}(\bm{\theta})=\\
&\underbrace{\sum_{X_{ij}\in\mathcal{X}_i}^{}\sum_{m=1}^{\vert \mathcal{N}(X_{ij})\vert}Q_{data}(X_{ij})\Big(c_{\bm{\theta}}\big(X_{ij}:\mathcal{N}\big)_m^2+2c_{\bm{\theta}}\big(X_{ij}:\mathcal{N}\big)_m\Big)}_{\Phi^{(i)}_1(\bm{\theta})}\notag\\
&\qquad-\underbrace{\sum_{X_{ij}\in\mathcal{X}_i}^{}\sum_{m=1}^{\vert\mathcal{N}(X_{ij})\vert}2\,Q_{data}(X_{im}) \,c_{\bm{\theta}}\big(X_{ijl}:\mathcal{N}\big)_m}_{\Phi^{(i)}_2(\bm{\theta})},\label{csm}
\end{align}
Further, considering all k groups, the optimizing criterion takes the form,
\begin{equation}
   \Phi_{CSM}(\bm{\theta})=\sum_{i=1}^{k}\Phi^{(i)}_{CSM}(\bm{\theta}).
\end{equation}
Here, $\vert\mathcal{N}(X_{ij})\vert$ denotes the size of neighbourhood of $X_{ij}.$  For large number of intervals, computation of $\eqref{csm}$ is time consuming.  Therefore, unbiased estimates of $\Phi_{CSM}({\bm{\theta}})$ is utilized. 

In $i$th group for $l$th device, the NOSD testing data point is denoted by a vector $Y_{il}$ of dimension $J_i+1$, which corresponds to $X_{ij}$ if failure occurs within the interval $(\tau_{i(j-1),\tau_{ij}}].$  Further, define reverse neighbourhood set $\mathcal{N}^{-1}(X^{'}_{ij})=\big\{(X_{ij},m)\vert \mathcal{N}(X_{ij})_m=X^{'}_{ij}\big\}$ where an element $(X_{ij},m)\in \mathcal{N}(X_{ij}^{'})$ indicates that $X^{'}_{ij}$ is the $m$th
neighbourhood of $X_{ij}.$  Based on these developments, the unbiased estimates of $\Phi^{(i)}_1$ and $\Phi^{(i)}_2$ extracted from the algorithms presented by Meng et al. \cite{meng2022concrete} is given as,
\begin{equation}
\begin{aligned}
\hat{\Phi}^{(i)}_1 &= \frac{1}{N_i}\sum_{l=1}^{N_i}\Bigg[\sum_{m=1}^{\vert \mathcal{N}(Y_{il})\vert}\Big(c_{\bm{\theta}}\big(Y_{il}:\mathcal{N}\big)_m^2 + 2\,c_{\bm{\theta}}\big(Y_{il}:\mathcal{N}\big)_m\Big)\Bigg], \\
\hat{\Phi}^{(i)}_2 &= \frac{1}{N_i}\sum_{l=1}^{N_i}\Bigg[\sum_{m=1}^{\vert\mathcal{N}^{-1}(Y^{'}_{il})\vert} 2\,c_{\bm{\theta}}\big(Y_{il}:\mathcal{N}\big)_m\Bigg],\label{escsm}
\end{aligned}
\end{equation} 
Thus we obtain,
\begin{equation}
    \hat{\Phi}^{(i)}_{CSM}({\bm{\theta}})=\hat{\Phi}^{(i)}_1-\hat{\Phi}^{(i)}_2,\quad
\text{and} \quad
   \hat{\Phi}_{CSM}({\bm{\theta}})=\sum_{i=1}^{k}\hat{\Phi}^{(i)}_{CSM},({\bm{\theta}}). \label{csmmax}
\end{equation}
The set of optimal tuning parameters $(\alpha,\beta,\gamma)$ can be obtained
by minimizing $\hat{\Phi}_{CSM}(\hat{\bm{\theta}}_{EP}).$  This optimization can be executed by an extensive grid search.

\section{Numerical study}\label{sec5}
This section carries the comparative study regarding optimal tuning parameters based on the proposed and existing methods discussed in the section \ref{sec4}.  The robustness of Exponential-polynomial divergence (EPD) based estimation over maximum likelihood estimate (MLE) is also discussed.

\subsection{\textbf{Simulation experiment}}\label{simsec}
For simulation analysis, 75 NOSD are divided into three groups and put under PSALT experimental setup.  The structure of simulated data is outlined in Table \ref{laysim}.  The count of failed NOSD units is recorded at each inspection time, and survived units are counted after termination time points $(0.7,0.8,0.5)$ for each group, respectively.

\begin{table}[htb!]
\caption{Layout for simulation experiment.}
\begin{center}
{\scalebox{1}{
\begin{tabular}{ccccccc}
\hline%
Groups&Devices&\multicolumn{3}{@{}c@{}}{Inspection times}&Stress rate\\
\hline
1&20&0.4&0.5&0.7&3\\
2&25&0.2&0.4&0.8&8\\
3&30&0.2&0.3&0.5&10\\
\hline
\end{tabular}}}
\end{center}
\label{laysim}
\end{table}
To study the robust behaviour of Minimum Exponential-polynomial 
divergence estimate (MEPDE), we need to examine how outliers affect estimates derived from the underlying model.  In this context, a contaminated version of data is taken by considering a contaminated model $U_{\epsilon}=(1-\epsilon)F(\bm{\theta})+\epsilon\,W(\bm{\theta}^{*}).$  Here, $\epsilon$ is the contamination rate which is taken as $\epsilon=(0,0.04,0.08,0.12,0.16)$ reflecting increasing levels of contamination.  By varying contamination rate $\epsilon$ the study can assess the sensitivity of the MEPDE to different levels of contamination and compare its performance with MLE.  The lifetime of NOSD data is generated from Log-logistic lifetime distribution $F(\bm{\theta})$ with the true model parameters set as $a=1.6$, $b=1.1$ and $\mu=2.7.$  The outliers are drawn from Weibull distribution $W(\bm{\theta}^{*})$ with parametric values specified as $(1.4,1.0,2.6).$

The optimal value of tuning parameters $(\alpha,\beta,\gamma)$ is determined through the proposed concrete score matching (CSM) method.   In this approach, the MEPDE of the parameters is obtained by implementing a coordinate-descent algorithm \cite{baghel2024robust,baghel2024analysis}.  The previously established methods IWJ, minAMAX, minMAE and minAMED are also employed to obtain the tuning parameters.  For comparative analysis, the sum of the root mean square (RMSE+) and absolute bias (Abs. bias) of the estimates is calculated using a Monte Carlo simulation of 1000 generations.  The optimal tuning parameters with corresponding RMSE+ values under the concerned methods are reported in Table \ref{tabtune}.

\begin{table}[htb!]
\caption{Set of optimal tuning parameters.}
\begin{center}
{\scalebox{0.8}{
\begin{tabular}{llcccccc}
\hline
\multirow{4}{*}{$(\alpha,\beta,\gamma)$}&0.00&(-7.5,0.1,0.74) &(-8.0,0.2,0.36) &(-6.0,0.6,0.12) & (-2.5,0.1,0.28)&(-1.0,0.2,1.0)\\
&0.04&(4.0,0.1,0.24)&(4.0,0.1,0.22)&(2.0,0.6,0.52)&(-1.0,0.8,0.94)&(-2.0,0.1,0.53)\\
&0.08&(2.5,0.8,0.48)&(-6.0,0.2,0.06)&(-2.5,0.6,0.86)&(-1.5,0.6,1.0)&(2.0,0.2,0.54)\\
&0.12&(2.0,1.0,1.0) & (-6.0,0.4,0.3)&(6.0,0.4,0.62)&(4.0,0.4,0.08)&(-7.5,0.3,0.1)\\
&0.16&(2.5,0.5,0.14) & (-6.5,0.1,0.3)&(6.0,0.2,0.2)&(-7.5,0.3,0.1)&(-7.5,0.3,0.1)\\
\hline
\multirow{4}{*}{RMSE+}&0.00&0.153449&0.219408&0.162074&0.270294&\textbf{0.109885}\\
&0.04&0.269560&0.275118&0.137002&\textbf{0.078686}&0.193437\\
&0.08&0.141004&0.338165&0.108285&\textbf{0.101061}&0.218728\\
&0.12&\textbf{0.126283}&0.269985&0.248003&0.363674&0.362706\\
&0.16&\textbf{0.338221}&0.399170&0.429380&0.397445&0.397445\\
\hline
\end{tabular}}}
\end{center}
\label{tabtune}
\end{table}

The table reveals that the estimates corresponding to $(\alpha,\beta,\gamma)$ obtained from existing methods perform quite efficiently for pure data and low contamination levels.  However, under high contamination ($\epsilon{=}0.12, 0.16$), estimates corresponding to $(\alpha,\beta,\gamma)$ obtained through proposed CSM method demonstrates greater efficiency compared to the existing approaches.  It is to be noted that at $\epsilon{=}0,$ the most efficient optimal tuning parameters are $(\alpha{=}-1.0,\beta{=}0.2,\gamma{=}1.0).$  Here, $\beta{=}0.2$ indicates that lower degree of contribution of the Bregman exponential divergence (BED) relative to the density power divergence (DPD) is sufficient for efficient estimation.  As the contamination increases to $(\epsilon{=}0.04,0.08),$ the most efficient optimal parameters shift to $(\alpha{=}-1.0,\beta{=}0.8,\gamma{=}0.94)$ and $(\alpha{=}-1.5,\beta{=}0.6,\gamma{=}1.0),$ respectively; here, higher values of $(\beta{=}0.8, 0.6)$ indicate an increased reliance on BED relative to DPD during the estimation process. 
 Moreover, at a contamination level of $\epsilon{=}0.12,$ the most efficient optimal parameters becomes $(\alpha{=}2.0,\beta{=}1.0,\gamma{=}1.0),$ where, $\beta{=}1$ signifies that the estimator relies solely on the BED component, rendering any choice of $\gamma$ redundant.  Finally, at $\epsilon{=}0.16,$ the most efficient optimal tuning parameters are $(\alpha{=}2.5,\beta{=}0.5,\gamma{=}0.14).$  Here $\beta=0.5,$ underscores the balanced integration of both divergence measures. 
 This behavior highlights the flexibility of the EPD measure, as it incorporates both DPD and BED as its special cases, adapting to varying contamination levels to achieve optimal estimation efficiency.

\begin{figure}[htb!]
\begin{center}
\subfloat[$\hat{a}$\label{bis}]{\includegraphics[height=4.5cm,width =0.33\textwidth]{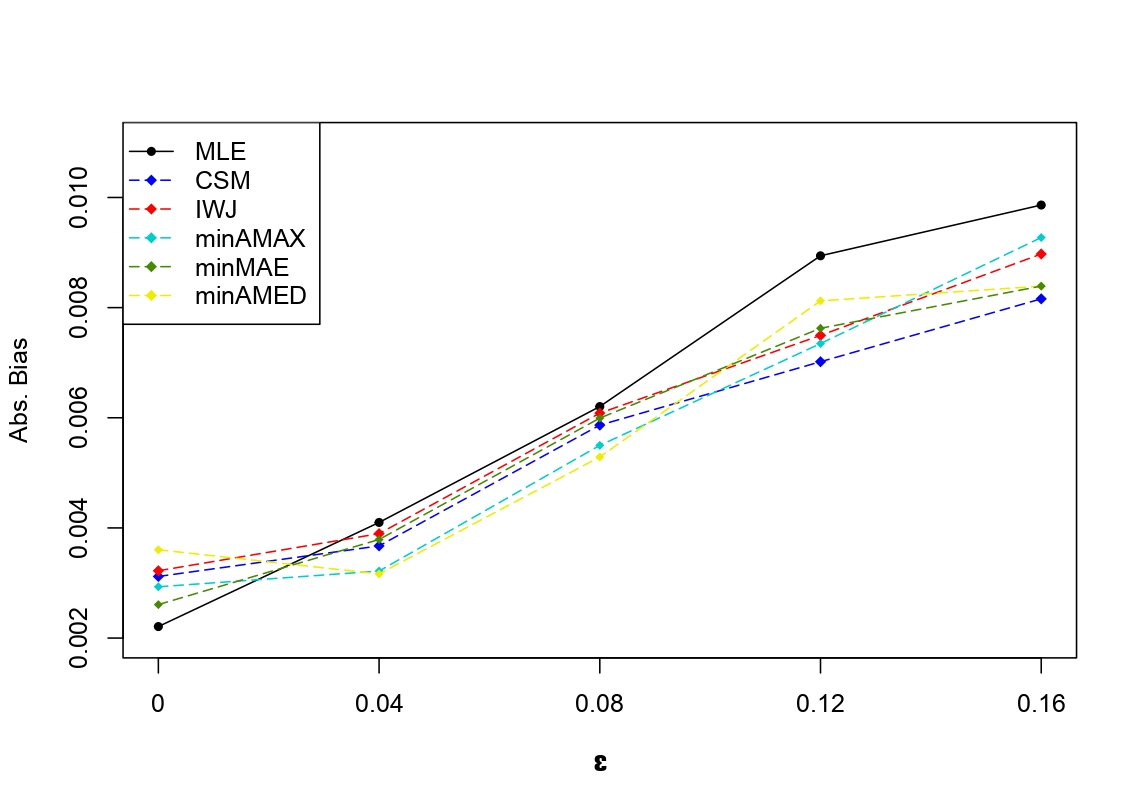}} 
\subfloat[$\hat{b}$]{\includegraphics[height=4.5cm,width =0.33\textwidth]{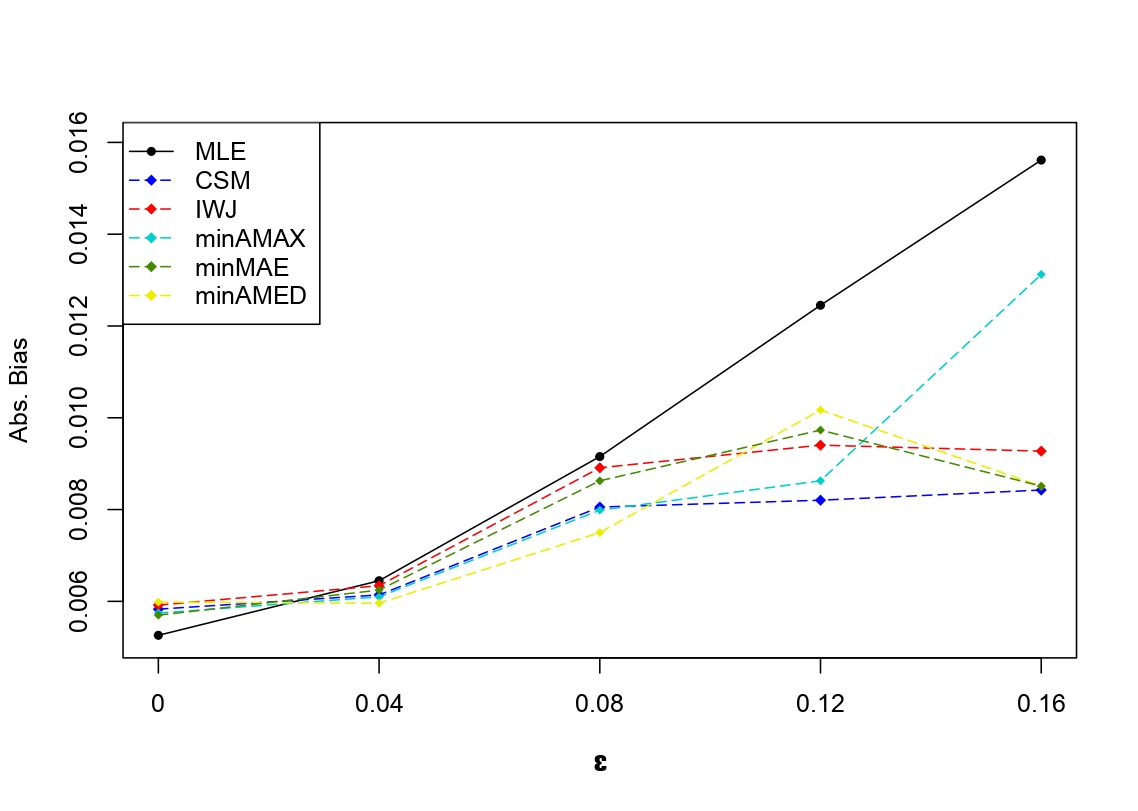}}
\subfloat[$\hat{\mu}$]
{\includegraphics[height=4.5cm,width =0.33\textwidth]{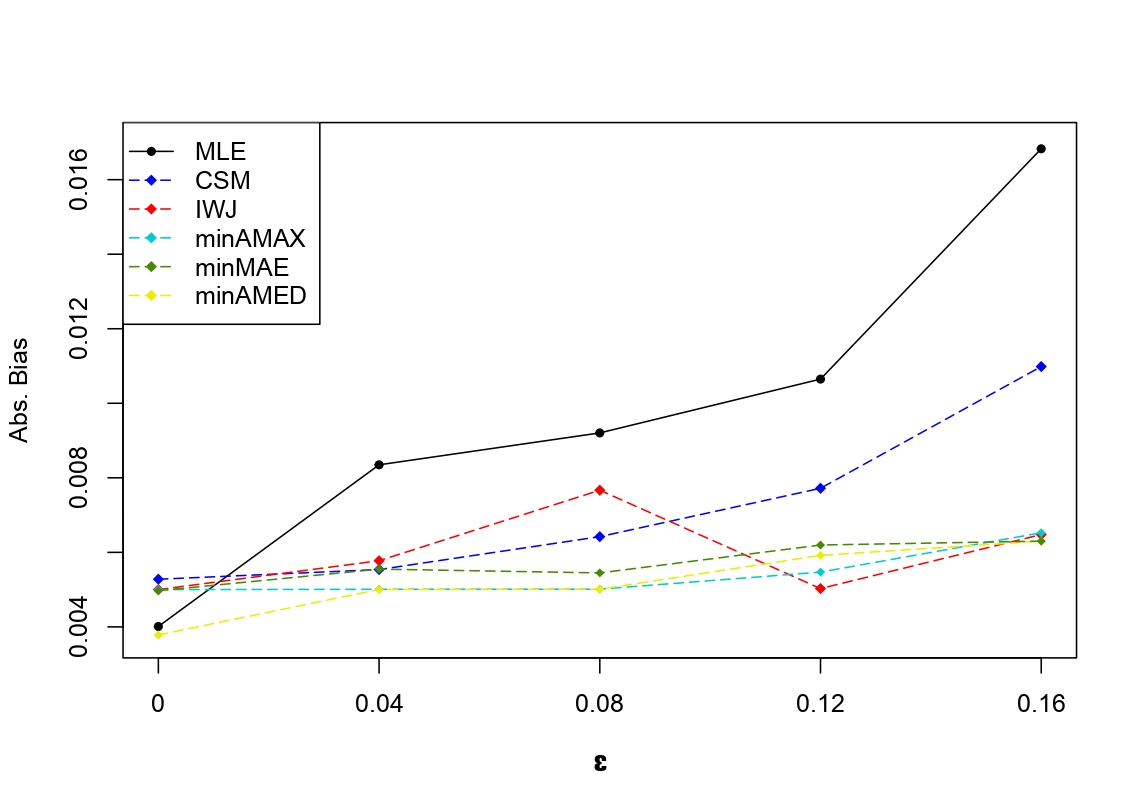}}
\end{center}
\caption{Absolute bias of parameters at different contamination rates.}
\label{bias}
\end{figure}

Furthermore, the robustness behaviour of the MEPDE, in comparison to the MLE, is examined through Figure \ref{bias}.  From this Figure, it is observed that when data is pure ($\epsilon=0$), the bias of MLE is the least.  As the contamination rate increases, the MEPDE estimates begin to outperform the MLE in terms of bias.  It is also observed that with the incorporation of contamination, the increase in absolute bias for the MEPDE is smaller than that of the MLE.  Thus, it is evident that MEPDE is a robust estimation method.

\subsection{\textbf{Data analysis}}
An analysis is carried out on the dataset determining the reliability of miniature light bulbs under a ramp-voltage experimental study conducted by Zhu \cite{zhu2010optimal}.  At the first stress level of 2.01 V/h, 62 light bulbs were used, while at the second stress level of 2.015 V/h, 61 light bulbs were used.  Hussam et al. \cite{hussam2022single} utilized this data for PSALT analysis.  For computational convenience in the present study, stress rates and failure times are multiplied by 0.1 and 0.01, respectively.  Therefore, the scaled stress rates $(\nu_1,\nu_2)$ and failure times used in this study are given in the table \ref{data}.  

\begin{table}[htb!]
\caption{Scaled failure times of miniature lightbulb data \cite{zhu2010optimal}.}
\begin{center}
{\scalebox{0.7}{
\begin{tabular}{ccccccccccc}
\hline
0.1357 &0.1992& 0.2330& 0.2781&0.3116& 0.3156 &0.3400& 0.4626&0.4641&0.5060&0.5676\\
0.5685&0.6013& 0.6500& 0.6586& 0.6620&0.6640& 0.6680& 0.6693&0.6825&0.7023& 0.7233\\
0.7260& 0.7543&0.7585& 0.7620& 0.7778& 0.7913 &0.8065& 0.8265&0.9033& 0.1451& 0.1561\\
0.1585& 0.1773 &0.1965&0.2105& 0.2120 &0.2421& 0.2485&0.3118& 0.3508& 0.4206& 0.4788\\
0.5421 &0.5455& 0.5585& 0.5643&0.5886& 0.6060&0.6248& 0.6281& 0.6341& 0.6376 &0.6418 \\
0.6615& 0.6641& 0.6991& 0.7173& 0.7246&0.7378 &0.7891\\
\hline
\multicolumn{10}{@{}c@{}}{$\nu_2=0.2015$}\\
\hline
0.1930& 0.2328& 0.2350 &0.2650& 0.2742& 0.2832& 0.2862& 0.3062& 0.3442& 0.3530 &0.3548\\
0.3830&0.4052& 0.4383& 0.4300& 0.4300& 0.4312& 0.4443& 0.4532 &0.4758& 0.4765& 0.4965\\
0.5142& 0.5127&0.5325& 0.5425& 0.5547 &0.5683& 0.5617& 0.0885& 0.1131& 0.1183& 0.1450\\
0.1483& 0.1773 &0.1935&0.2550& 0.2615& 0.2745& 0.2761& 0.2805& 0.3096& 0.3100& 0.3481\\
0.3603& 0.4308& 0.4563& 0.4603& 0.4633 &0.4962& 0.4986 &0.5066& 0.5093& 0.5103& 0.5173\\
0.5195 &0.5236& 0.5478& 0.5558& 0.5583&0.5713\\
\hline
\end{tabular}}}
\end{center}
\label{data}
\end{table}

For inferential analysis, failures are recorded at the inspection times $\tau_1{=}(0.37,0.67,0.75)$ and $\tau_2{=}(0.37,0.44,0.54)$ for the two progressive stress levels, respectively.  To assess whether the Log-logistic distribution fits the data for the given model, a bootstrap-based goodness-of-fit test is conducted, and an approximated p-value is computed.  The test statistic used for conducting the test is defined as the distance-based measure, $$TS=\sum_{i=1}^{k}\sum_{j=1}^{J_i+1}\left\vert\frac{n_{ij}-\hat{n}_{ij}}{\hat{n}_{ij}}\right\vert,$$
where, $n_{ij}$ represents observed and $\hat{n}_{ij}$ denotes the expected failure/survival counts.  The calculated value of the test statistic is $TS=3.848737$, and the corresponding approximate p-value is 0.756, indicating strong evidence for the suitability of the distribution of the data.

\begin{table}[htb!]
\caption{Set of optimal tuning parameters for data.}
\begin{center}
{\scalebox{1}{
\begin{tabular}{lcc}
\hline%
Methods&$(\alpha,\beta,\gamma)$&BT RMSE$^{+}$\\
\hline
CSM&( -15,1.0,0.02)&0.002379\\
IWJ&(-6,0.1,0.16)&\textbf{0.001378}\\
minAMAX&(-6,1.0,0.02)&0.009045\\
minMAE&(-6,1.0,0.02)&0.009045\\
minAMED&(9,1.0,0.02)&0.001907\\
\hline
\end{tabular}}}
\end{center}
\label{optdata}
\end{table}

\begin{table}[htb!]
\caption{Parameter estimates, bootstrap bias (BT Bias) and 95$\%$ confidence limits for data.}
\begin{center}
{\scalebox{0.8}{
\begin{tabular}{lcccccc}
\hline%
\multirow{2}{*}{$(\alpha,\beta,\gamma)$}&\multicolumn{2}{@{}c@{}}{$\hat{a}$}&\multicolumn{2}{@{}c@{}}{$\hat{b}$}&\multicolumn{2}{@{}c@{}}{$\hat{\mu}$}\\\cmidrule(lr){2-3}\cmidrule(lr){4-5}\cmidrule(lr){6-7}
& Est.& BT Bias& Est.& BT Bias& Est.& BT Bias\\
\hline
MLE&3.197141&  0.002903&  0.123643& -0.013645&  3.200343 & 0.000237\\
CSM&3.199962 & \textbf{0.000037} & 0.107794 & 0.002204 & 3.200135 &-0.000135\\
IWJ&3.199172&  0.000828&  0.110429& -0.000429&  3.200058& -0.000051\\
minAMAX&3.159359&  0.000064&  0.110021& -0.000405 & 3.200015&  \textbf{0.000029}\\
minMAE&3.159359&  0.000064&  0.110021& -0.000405 & 3.200015&  \textbf{0.000029}\\
minAMED&3.199463&  0.000537&  0.110099 &\textbf{-0.000106} & 3.200012 & 0.000046\\
\hline
& LL&UL &LL & UL &LL & UL\\
\hline
MLE&3.139324& 3.254959& 0.116062& 0.131226& 3.061116& 3.339571\\
CSM&3.124180& 3.275730& 0.093744 &0.121843 &2.938250 &3.462019\\
IWJ&3.144369& 3.253975& 0.103906& 0.116953& 3.084548 &3.315568\\
minAMAX&3.094637& 3.224080& 0.099534& 0.120508& 3.021821& 3.378208\\
minMAE&3.094637& 3.224080& 0.099534& 0.120508& 3.021821& 3.378208\\
minAMED&3.127626& 3.271300& 0.099859& 0.120339& 3.030453 &3.369571\\
\hline
\end{tabular}}}
\end{center}
\label{oudata}
\end{table}

The optimal values of the tuning parameters $(\alpha,\beta,\gamma)$ for MEPDE are determined using the methods discussed earlier and are presented in Table \ref{optdata}.  The sum of the bootstrap RMSE (BT RMSE+) for the estimates of the parameters $(a,b,\mu)$ is determined to identify the best tuning method for this data.  This table shows that the MEPDE with optimal parameters $(\alpha{=}-6,\beta{=}0.1,\gamma{=}0.16)$ obtained through IWJ results in the lowest BT RMSE+ values due to absence of a high percentage of outliers in data.  Otherwise, MEPDE with optimal parameters determined through proposed CSM method would have been more efficient.  From this table, it should be noted that $\beta{=}1$ signifies that the estimator relies solely on the BED component, rendering any choice of $\gamma$ redundant.  Furthermore, The estimates derived from MLE and MEPDE with $95\%$ asymptotic confidence intervals (Lower Limit: LL and Upper Limit: UL) and bootstrap bias (BT Bias) are reported in table \ref{oudata}.  From this table, it is observed that BT Bias of MLE is more than that of MEPDE demonstrating the preference of MEPDE over MLE.

\section{Optimal design}\label{sec6}
This section focuses on the optimal strategy for determining the sample allocation $N_i$ and inspection times $\tau_{ij};i{=}1,2,{\dots},k; j{=}1,2,{\dots},J_i$ through minimizing the trace of the asymptotic variance-covariance matrix of the estimators.  First, let us assume that the total experimental cost $C$ consists of the following components.
\begin{equation}
C=c_a+c_u\sum_{i=1}^{k}N_i+c_0\sum_{i=1}^{k}\tau_{iJ_i}+c_s\sum_{i=1}^{k}J_i-(N-D)\,c_v.
\end{equation}
where $c_a$, $c_u$, $c_0$, and $c_s$ represent user-defined values corresponding to the installation cost of the experiment, sample cost per item, operation cost per unit of time and inspection cost per item, respectively.  Here, $c_v$ is salvage value per survived item at termination time such that $0{\leq}c_v{<}c_u$, $D$ is expected total number of failures obtained as $D{=}E\left(\sum_{i=1}^{k}\sum_{j=1}^{J_i}N_{ij}\right).$   The component $(N-D)\,c_v$ is frequently used as part of cost constraints in various optimal design studies \cite{bhattacharya2014optimum,bhattacharya2017computation,sen2018statistical,sen2021statistical}.
Further, it is assumed that total experimental cost C can not exceed certain budget say $C_r,$ i.e. 
\begin{equation}
C \leq C_r. \label{cost}
\end{equation}

 The objective function is defined as the trace of asymptotic variance-covariance matrix of the estimates, $Tr(V(\hat{\bm{\theta}})).$  Therefore, optimization problem is formulated as
\begin{equation}
\begin{aligned}
&\Phi=\min Tr(V(\hat{\bm{\theta}})),\\
&\text{Subject to}\\
&C\leq C_r\;;\;N_i>0\;;\;i=1,\dots,k.\\
&\tau_{i1}\leq\tau_{i2}\leq\dots\leq\tau_{iJ_i}\;;\;i=1,\dots,k;\\
&\max\limits_{1,2,\dots,k}(\tau_{iJ_i})\leq\tau_{max.}.\label{opjfn}
\end{aligned}
\end{equation}
where, $\tau_{max}$ denotes maximum allowable termination time of experiment.  Dealing with large sample sizes and multiple stress levels under constraints presents significant analytical and computational challenges. 
 While there are various methods for solving constrained optimization problems, this study explores an approach by employing the constrained particle swarm optimization (CPSO) algorithm.

 \subsection{\textbf{Constraint particle swarm optimization (CPSO)}}
Particle Swarm Optimization (PSO), introduced by Kennedy and Eberhart \cite{kennedy1995particle}, is a meta-heuristic algorithm inspired by the social behaviour of swarms \cite{hassan2004particle,kim2010simple}. 
 Readers may refer to \cite{li2021effective,yeh2010particle,xiahou2022fusing} and references cited therein for the utility of PSO in various scenarios.  PSO solves optimization problems by simulating the movement of particles in a search space, where each particle represents a potential solution.  Each particle in the swarm is characterized by its position and velocity, which are updated iteratively to explore the search space.  The updates are guided by two key components: Personal Best (Pbest), the best position a particle has found, and Global Best (Gbest), the best position found by any particle in the swarm.  

While PSO was initially designed for unconstrained optimization problems, the inevitability of constraints in real-world applications motivated researchers to develop constraint-handling techniques for PSO \cite{hu2002solving,innocente2021constraint}.  For the present study, we adopt an approach inspired by Ang et al. \cite{ang2020constrained} to enable particles to navigate the search space effectively while satisfying problem-specific constraints.  The steps of Constraint Particle Swarm Optimization (CPSO) are outlined in Algorithm \ref{cpso}, where \( X \) represents the solution set of a particle, specifically defined for the present study as follows.
\[
X = \left\{
\begin{array}{cccccc}
N_1\;;&\tau_{11}&\dots&\tau_{1j}&\dots&\tau_{1J_1}\\
N_2\;;&\tau_{21}&\dots&\tau_{2j}&\dots&\tau_{2J_2}\\
\vdots & \vdots & \vdots & \vdots& \vdots & \vdots \\
N_i\;;&\tau_{i1}&\dots&\tau_{ij}&\dots&\tau_{iJ_i}\\
\vdots & \vdots & \vdots & \vdots& \vdots & \vdots \\
N_k\;;&\tau_{k1}&\dots&\tau_{kj}&\dots&\tau_{kJ_k}\\
\end{array}
\right\}.
\]

\begin{algorithm}[htb!]
\caption{{\textbf{Constraint particle swarm optimization (CPSO)}}}\label{cpso}
\begin{itemize}
\item Initialize a feasible swarm of $L$ particles having initial solution set $(X^{(0)}_1,\dots,X^{(0)}_L)$ and set of initial velocities $V^{(0)}_l=0\;;\; l=1,\dots,L.$
\item Calculate the $\Phi^{(0)}(X_l)$ for each of the particles and find $Gbest^{(0)}.$
\item[]For $t=t+1$, 
\item Update velocity for each particle using the following rule, 
\begin{equation}
\begin{aligned}
V_{l}^{(t+1)}&= w\,V_{l}^{(t)} + c_1 r^{(t)}_1 (Pbest^{(t)}_l - X_{l}^{(t)}) \notag\\
&\qquad+c_2 r^{(t)}_2 (Gbest - X_{l}^{(t)})\\
X_{l}^{(t+1)}&= X_{l}^{(t)} + V_{l}^{(t+1)}
\end{aligned},\label{psoeq}
\end{equation}
where, \( V_{l}^{(t)}\) and \( X_{l}^{(t)}\) is the velocity and position of the \(l\)th particle at \(t\)th iteration, respectively, \( w\) is inertia weight, \( c_1,c_2\) are acceleration coefficients and \( r_1,r_2\) are random numbers in $[0,1].$ 
\item Find $Pbest^{(t+1)}_l$ through Deb's rule given in Algorithm \ref{algdeb}.
\item Find global best position as, 
$
Gbest^{(t+1)}=\min\limits_{l=1,\dots,L}\,\Phi(Pbest^{(t+1)}_l).
$
\item Continue the iteration until a stopping criterion is satisfied.
\end{itemize}
\end{algorithm}

If constraints $N_i>0$ and $\tau_{ij}\leq \tau_{i(j+1)}$ are violated, we keep updating velocity using equation \eqref{psoeq} until particle comes back to feasible region \cite{innocente2021constraint}.  Further, Deb's rule \cite{deb2000efficient} is adopted according to the guidelines given by Ang et al. \cite{ang2020constrained} to handle other constraints.  Deb's rule for updating \textbf{Pbest} is presented in the Algorithm \ref{algdeb}.  Let $\psi$ denote degree of constraint violation for a solution set X,
\begin{align}
    \psi(X)&=\left[\max\bigg\{0,(C-C_r)\bigg\}\right.\left.+\sum_{i=1}^{k}\sum_{j=1}^{J_i}\max\bigg\{0,\Big(\max(\tau_{iJ_i})-\tau_{max}\Big)\bigg\}\right].\label{constvio}
\end{align}

\begin{algorithm}[htb!]
\caption{{\textbf{Deb's rule}}}\label{algdeb}
\begin{itemize}
\item [] For any $l$th particle at $(t+1)$th iteration,
\item If $\psi(X^{(t+1)}_l)\leq 0$ and $\psi(Pbest^{(t)}_l)\leq 0$ (If both solutions are feasible),
$$Pbest^{(t+1)}_l = 
\begin{cases}
Pbest^{(t)}_l \quad \text{if} \ \Phi(X^{(t+1)}_l)>\Phi(Pbest_{l}^{(t)})\\
X^{(t+1)}_l \;\;\quad \text{if}\ \Phi(X^{(t+1)}_l)\leq \Phi(Pbest_{l}^{(t)}) \\
\end{cases}.
$$
\item If $\psi(X^{(t+1)}_l)\leq 0$ and $\psi(Pbest^{(t)}_l)> 0,$ $Pbest^{(t+1)}_l=X^{(t+1)}_l.$
\item[] If $\psi(X^{(t+1)}_l)> 0$ and $\psi(Pbest^{(t)}_l)\leq 0,$ $Pbest^{(t+1)}_l=Pbest^{(t)}_l.$
\item If $\psi(X^{(t+1)}_l)> 0$ and $\psi(Pbest^{(t)}_l)> 0$ (If both solutions are infeasible),
$$Pbest^{(t+1)}_l = 
\begin{cases}
Pbest^{(t)}_l \quad \text{if} \ \psi(X^{(t+1)}_l)>\psi(Pbest_{l}^{(t)})\\
X^{(t+1)}_l \;\;\quad \text{if}\ \psi(X^{(t+1)}_l)\leq \psi(Pbest_{l}^{(t)}) \\
\end{cases}.
$$
\end{itemize}
\end{algorithm}

\begin{table}[htb!]
\caption{Optimal Design when $\beta=0$.}
\begin{center}
{\scalebox{0.7}{
\begin{tabular}{ccccccccc}
\hline%
$(\gamma)$&$N_{opt}$&\multicolumn{3}{@{}c@{}}{$\tau_{opt}$}&$C_{opt}$&$Tr[V(\hat{\bm{\theta}})]$\\
\cmidrule(lr){1-1}\cmidrule(lr){2-2}\cmidrule(lr){3-5}\cmidrule(lr){6-6}\cmidrule(lr){7-7}
\multirow{3}{*}{$(0.3)$}&06 & 0.567& 0.577& 0.810 &\multirow{3}{*}{8150.890}&\multirow{3}{*}{ \textbf{0.000071}}\\
&35 &  0.093& 0.324& 0.350 \\
&25 & 0.182& 0.297& 0.369 \\
\hline
\multirow{3}{*}{$(0.6)$}&40 &  0.212& 0.551 &0.558&\multirow{3}{*}{9235.490}&\multirow{3}{*}{0.005618}\\
&32 & 0.064& 0.505& 0.544 \\
&03 &  0.104& 0.262& 0.709 \\
\hline
\multirow{3}{*}{$(0.9)$}&18 & 0.066& 0.210& 0.732 &\multirow{3}{*}{ 6352.780 }&\multirow{3}{*}{0.003417}\\
&18 & 0.130& 0.472& 0.823 \\
&15 & 0.240& 0.314& 0.440 \\
\hline
\end{tabular}}}
\end{center}
\label{b0}
\end{table}

\begin{table}[htb!]
\caption{Optimal Design when $\beta=0.2$.}
\begin{center}
{\scalebox{0.6}{
\begin{tabular}{ccccccccc}
\hline%
$(\alpha,\gamma)$&$N_{opt}$&\multicolumn{3}{@{}c@{}}{$\tau_{opt}$}&$C_{opt}$&$Tr[V(\hat{\bm{\theta}})]$\\
\cmidrule(lr){1-1}\cmidrule(lr){2-2}\cmidrule(lr){3-5}\cmidrule(lr){6-6}\cmidrule(lr){7-7}
\multirow{3}{*}{$(-6,0.3)$}&38 & 0.202& 0.514& 0.545&\multirow{3}{*}{ 8512.575}&\multirow{3}{*}{0.002042}\\
&30 & 0.082& 0.475& 0.516\\
& 01& 0.116& 0.276& 0.713 \\
\hline
&40 &  0.212& 0.551& 0.558 &\multirow{3}{*}{9239.490 }&\multirow{3}{*}{0.005619}\\
$(-6,0.6)$&32 &0.064& 0.505& 0.544 \\
&03 & 0.104& 0.262& 0.709\\
\hline
\multirow{3}{*}{$(-6,0.9)$}&18 & 0.060& 0.196& 0.729 &\multirow{3}{*}{ 6470.735}&\multirow{3}{*}{0.003470}\\
&18 & 0.119& 0.467 &0.831 \\
&16 & 0.241& 0.314 &0.442 \\
\hline
\multirow{3}{*}{$(-4,0.3)$}&39 & 0.253 &0.526& 0.725 &\multirow{3}{*}{ 8765.650 }&\multirow{3}{*}{\textbf{0.001662}}\\
&20 &0.149& 0.322& 0.337\\
&12 & 0.202& 0.382& 0.710 \\
\hline
&40 &  0.212& 0.551& 0.558 &\multirow{3}{*}{9239.490 }&\multirow{3}{*}{0.005619}\\
$(-4,0.6)$&32 &0.064& 0.505& 0.544 \\
&03 & 0.104& 0.262& 0.709\\
\hline
\multirow{3}{*}{$(-4,0.9)$}&03 & 0.078& 0.238& 0.736 &\multirow{3}{*}{ 4800.595}&\multirow{3}{*}{0.005670}\\
&22 & 0.143& 0.476& 0.814\\
&13 & 0.247& 0.315& 0.454 \\
\hline
\multirow{3}{*}{$(-2,0.3)$}&36 & 0.196& 0.500& 0.533&\multirow{3}{*}{8277.840 }&\multirow{3}{*}{0.002068 }\\
&30 & 0.086& 0.498& 0.561 \\
&01 &0.127& 0.290& 0.718\\
\hline
&40 &  0.212& 0.551& 0.558 &\multirow{3}{*}{9239.490 }&\multirow{3}{*}{0.005619}\\
$(-2,0.6)$&32 &0.064& 0.505& 0.544 \\
&03 & 0.104& 0.262& 0.709\\
\hline
\multirow{3}{*}{$(-2,0.9)$}&22 &0.074 &0.230& 0.738 &\multirow{3}{*}{4200.990}&\multirow{3}{*}{0.006035}\\
&10 & 0.140& 0.473& 0.816 \\
&01 & 0.247& 0.315& 0.456 \\
\hline%
\multirow{3}{*}{$(2,0.3)$}&40 & 0.212& 0.551& 0.558 &\multirow{3}{*}{9235.490 }&\multirow{3}{*}{0.004014}\\
&32 & 0.064& 0.505& 0.544 \\
&03 & 0.104& 0.262& 0.709  \\
\hline
\multirow{3}{*}{$(2,0.6)$}& 38& 0.207& 0.534 &0.549 &\multirow{3}{*}{8518.930 }&\multirow{3}{*}{0.004012}\\
&30 & 0.085& 0.498& 0.560\\
&01 &  0.130& 0.294& 0.721
  \\
\hline
\multirow{3}{*}{$(2,0.8)$}&24 & 0.076& 0.236& 0.739 &\multirow{3}{*}{6713.595 }&\multirow{3}{*}{0.003729}\\
&15 & 0.145& 0.479& 0.813\\
&15 & 0.242& 0.314 &0.445 \\
\hline
\multirow{3}{*}{$(4,0.3)$}&40 &0.212& 0.551& 0.558 &\multirow{3}{*}{9239.490 }&\multirow{3}{*}{0.004216}\\
&32 &0.064& 0.505& 0.544\\
&03 & 0.104 &0.262& 0.709  \\
\hline
\multirow{3}{*}{$(4,0.6)$}&38 & 0.194& 0.496& 0.530 &\multirow{3}{*}{ 8519.095 }&\multirow{3}{*}{0.002919}\\
&30 & 0.093& 0.496& 0.567 \\
&01 & 0.167& 0.337& 0.736  \\
\hline
\multirow{3}{*}{$(4,0.9)$}&23 & 0.063& 0.202 &0.727&\multirow{3}{*}{  7674.670}&\multirow{3}{*}{0.005629}\\
&24 &  0.145& 0.477& 0.812 \\
&15 & 0.244& 0.315 &0.448 \\
\hline
\multirow{3}{*}{$(6,0.3)$}& 22& 0.139& 0.387& 0.688&\multirow{3}{*}{5882.415}&\multirow{3}{*}{{0.002552}}\\
&24 & 0.173& 0.512& 0.687\\
&01 & 0.173& 0.254& 0.457  \\
\hline
\multirow{3}{*}{$(6,0.6)$}&40 & 0.212& 0.551& 0.558 &\multirow{3}{*}{9240.490 }&\multirow{3}{*}{0.006610}\\
&32 & 0.064& 0.505& 0.544 \\
&03 & 0.104& 0.262& 0.709  \\
\hline
\multirow{3}{*}{$(6,0.8)$}&39 &0.192& 0.506& 0.560 &\multirow{3}{*}{8397.205}&\multirow{3}{*}{0.006963}\\
&28 & 0.090& 0.480& 0.585 \\
&01 & 0.165& 0.345& 0.745 \\
\hline
\end{tabular}}}
\end{center}
\label{b22}
\end{table}

\begin{table}[htb!]
\caption{Optimal Design when $\beta=0.5$.}
\begin{center}
{\scalebox{0.6}{
\begin{tabular}{ccccccccc}
\hline%
$(\alpha,\gamma)$&$N_{opt}$&\multicolumn{3}{@{}c@{}}{$\tau_{opt}$}&$C_{opt}$&$Tr[V(\hat{\bm{\theta}})]$\\
\cmidrule(lr){1-1}\cmidrule(lr){2-2}\cmidrule(lr){3-5}\cmidrule(lr){6-6}\cmidrule(lr){7-7}
\multirow{3}{*}{$(-6,0.3)$}&36 & 0.196& 0.500& 0.533 &\multirow{3}{*}{8277.840}&\multirow{3}{*}{{0.002073}}\\
&30 & 0.086& 0.498& 0.561 \\
&01 & 0.127& 0.290& 0.718 \\
\hline
& 40& 0.212& 0.551& 0.558 &\multirow{3}{*}{ 9239.490 }&\multirow{3}{*}{0.005618}\\
$(-6,0.6)$&32 &  0.064& 0.505& 0.544 \\
&03 & 0.104& 0.262& 0.709  \\
\hline
\multirow{3}{*}{$(-6,0.9)$}&28 & 0.047& 0.159 &0.718 &\multirow{3}{*}{ 9230.535}&\multirow{3}{*}{0.009697}\\
&28 & 0.107& 0.462& 0.839 \\
&19 &0.261& 0.316& 0.474  \\
\hline
\multirow{3}{*}{$(-4,0.3)$}&18 & 0.177& 0.386& 0.697 &\multirow{3}{*}{ 5889.175}&\multirow{3}{*}{0.002517}\\
&23 & 0.200& 0.515& 0.712 \\
&06 & 0.196& 0.284& 0.441  \\
\hline
& 40& 0.212& 0.551& 0.558 &\multirow{3}{*}{ 9239.490 }&\multirow{3}{*}{0.005618}\\
$(-4,0.6)$&32 &  0.064& 0.505& 0.544 \\
&03 & 0.104& 0.262& 0.709  \\
\hline
\multirow{3}{*}{$(-4,0.9)$}& 28& 0.047& 0.159& 0.718 &\multirow{3}{*}{ 9231.535}&\multirow{3}{*}{0.009843}\\
&28 & 0.107& 0.462& 0.839
\\
&19 & 0.261& 0.316& 0.474  \\
\hline
\multirow{3}{*}{$(-2,0.3)$}&40 & 0.212& 0.551& 0.558 &\multirow{3}{*}{9239.490}&\multirow{3}{*}{0.003962}\\
& 32& 0.064& 0.505 &0.544 \\
&03 & 0.104& 0.262& 0.709 \\
\hline
\multirow{3}{*}{$(-2,0.6)$}&40 & 0.212 &0.551& 0.558 &\multirow{3}{*}{9239.490}&\multirow{3}{*}{0.005925}\\
&32 &0.064& 0.505 &0.544 \\
&03 & 0.104& 0.262& 0.709  \\
\hline
\multirow{3}{*}{$(-2,0.9)$}&20 &0.066 &0.213& 0.732 &\multirow{3}{*}{ 6956.945}&\multirow{3}{*}{0.008424}\\
&25 & 0.128& 0.471& 0.825 \\
&11 & 0.255& 0.315& 0.466  \\
\hline
\multirow{3}{*}{$(2,0.3)$}&40 &  0.212& 0.551 &0.558 &\multirow{3}{*}{9239.490}&\multirow{3}{*}{0.004357 }\\
&32 & 0.064& 0.505& 0.544\\
&03 &0.104& 0.262& 0.709 \\
\hline
\multirow{3}{*}{$(2,0.6)$}&37 & 0.210& 0.524& 0.551 &\multirow{3}{*}{8045.600
 }&\multirow{3}{*}{0.005499 }\\
&27 & 0.083 &0.515 &0.600 \\
&01 &0.126& 0.281& 0.692 \\
\hline
\multirow{3}{*}{$(2,0.9)$}& 30& 0.183& 0.454& 0.516 &\multirow{3}{*}{7683.600}&\multirow{3}{*}{0.004974
 }\\
&29 & 0.141& 0.448& 0.545 \\
&03 & 0.235& 0.418& 0.757\\
\hline
\multirow{3}{*}{$(4,0.3)$}&37 & 0.180& 0.482& 0.530 &\multirow{3}{*}{6363.570
 }&\multirow{3}{*}{  0.002276}\\
&13 & 0.095& 0.466& 0.571 \\
&01 &  0.204& 0.426& 0.766\\
\hline
\multirow{3}{*}{$(4,0.6)$}&33 & 0.203& 0.530& 0.556 &\multirow{3}{*}{6966.160}&\multirow{3}{*}{ 0.004913}\\
&22 & 0.090& 0.500& 0.579 \\
&01 & 0.111& 0.282& 0.715 \\
\hline
\multirow{3}{*}{$(4,0.8)$}&36 & 0.211& 0.553& 0.588 &\multirow{3}{*}{8513.945 }&\multirow{3}{*}{0.001290 }\\
&29 & 0.086& 0.495& 0.598 \\
&04 & 0.095& 0.351& 0.767 \\
\hline
\multirow{3}{*}{$(6,0.3)$}&40 & 0.212& 0.551& 0.558 &\multirow{3}{*}{9240.490 }&\multirow{3}{*}{0.004739 }\\
&32 & 0.064& 0.505& 0.544\\
&03 & 0.104& 0.262& 0.709 \\
\hline
\multirow{3}{*}{$(6,0.6)$}&36 & 0.176& 0.495& 0.590 &\multirow{3}{*}{ 6605.205}&\multirow{3}{*}{\textbf{0.000185 }}\\
&15 & 0.086& 0.509& 0.632 \\
&02 &  0.110 &0.327& 0.731\\
\hline
\multirow{3}{*}{$(6,0.7)$}&11 & 0.168 &0.525 &0.626 &\multirow{3}{*}{1816.445 }&\multirow{3}{*}{0.020871 }\\
&01 & 0.077& 0.487 &0.567\\
&01 & 0.111& 0.287& 0.709 \\
\hline
\end{tabular}}}
\end{center}
\label{b52}
\end{table}

\begin{table}[htb!]
\caption{Optimal Design when $\beta=0.8$.}
\begin{center}
{\scalebox{0.6}{
\begin{tabular}{ccccccccc}
\hline%
$(\alpha,\gamma)$&$N_{opt}$&\multicolumn{3}{@{}c@{}}{$\tau_{opt}$}&$C_{opt}$&$Tr[V(\hat{\bm{\theta}})]$\\
\cmidrule(lr){1-1}\cmidrule(lr){2-2}\cmidrule(lr){3-5}\cmidrule(lr){6-6}\cmidrule(lr){7-7}
&40 & 0.212& 0.551& 0.558 &\multirow{3}{*}{ 9239.490}&\multirow{3}{*}{{0.003943 }}\\
$(-6,0.3)$&32 & 0.064& 0.505& 0.544\\
&03 &  0.104& 0.262& 0.709\\
\hline
&40 & 0.212& 0.551& 0.558 &\multirow{3}{*}{ 9239.490}&\multirow{3}{*}{0.005608 }\\
$(-6,0.6)$&32 & 0.064 &0.505& 0.544 \\
&03 & 0.104& 0.262& 0.709\\
\hline
\multirow{3}{*}{$(-6,0.9)$}&28 & 0.047& 0.159& 0.718 &\multirow{3}{*}{9232.535 }&\multirow{3}{*}{ 0.010586}\\
&28 & 0.107& 0.462 &0.839 \\
&19 & 0.261& 0.316& 0.474 \\
\hline
&40 & 0.212& 0.551& 0.558 &\multirow{3}{*}{ 9239.490}&\multirow{3}{*}{{0.003943 }}\\
$(-4,0.3)$&32 & 0.064& 0.505& 0.544\\
&03 &  0.104& 0.262& 0.709\\
\hline
&40 & 0.212& 0.551& 0.558 &\multirow{3}{*}{ 9239.490}&\multirow{3}{*}{0.005608 }\\
$(-4,0.6)$&32 & 0.064 &0.505& 0.544 \\
&03 & 0.104& 0.262& 0.709\\
\hline
\multirow{3}{*}{$(-4,0.9)$}& 30& 0.101& 0.205& 0.386 &\multirow{3}{*}{9230.400}&\multirow{3}{*}{0.004932 }\\
&41 & 0.357& 0.411& 0.765 \\
&04 & 0.441& 0.664& 0.853\\
\hline
\multirow{3}{*}{$(-2,0.3)$}&40 & 0.212& 0.551& 0.558 &\multirow{3}{*}{9239.490}&\multirow{3}{*}{0.004147 }\\
&32 & 0.064& 0.505 &0.544 \\
&03 & 0.104& 0.262& 0.709 \\
\hline
\multirow{3}{*}{$(-2,0.6)$}&40 & 0.212& 0.551& 0.558 &\multirow{3}{*}{ 9239.490 }&\multirow{3}{*}{0.006661 }\\
&32 &  0.064 &0.505& 0.544\\
&03 & 0.104& 0.262& 0.709 \\
\hline
\multirow{3}{*}{$(-2,0.9)$}&27 & 0.048& 0.167& 0.720
 &\multirow{3}{*}{8628.920 }&\multirow{3}{*}{0.010228 }\\
&27 & 0.112& 0.461& 0.834 \\
& 16& 0.257& 0.316& 0.469\\
\hline
\multirow{3}{*}{$(2,0.3)$}&40 & 0.212& 0.551& 0.558 &\multirow{3}{*}{9237.490 }&\multirow{3}{*}{ 0.005147}\\
&32 & 0.064& 0.505& 0.544 \\
&03 &  0.104& 0.262 &0.709 \\
\hline
\multirow{3}{*}{$(2,0.6)$}&40 & 0.212& 0.551& 0.558
 &\multirow{3}{*}{ 9236.490}&\multirow{3}{*}{ 0.008124 }\\
&32 & 0.064& 0.505& 0.544 \\
&03 & 0.104& 0.262 &0.709\\
\hline
\multirow{3}{*}{$(2,0.9)$}&40 & 0.212& 0.551& 0.558
 &\multirow{3}{*}{  9243.490}&\multirow{3}{*}{ 0.011929 }\\
&32 & 0.064& 0.505& 0.544 \\
&03 & 0.104& 0.262 &0.709\\
\hline
\multirow{3}{*}{$(4,0.3)$}&37 & 0.183& 0.463 &0.519 &\multirow{3}{*}{ 8640.515 }&\multirow{3}{*}{  0.003248}\\
&31 & 0.132& 0.481& 0.598 \\
&02 & 0.239& 0.429& 0.773 \\
\hline
\multirow{3}{*}{$(4,0.6)$}& 30&  0.101& 0.205 &0.386&\multirow{3}{*}{9230.400}&\multirow{3}{*}{0.000827 }\\
&41 & 0.357 &0.411 &0.765 \\
&04 &  0.441& 0.664& 0.853 \\
\hline
\multirow{3}{*}{$(4,0.9)$}&40 &  0.212& 0.551& 0.558&\multirow{3}{*}{9232.490}&\multirow{3}{*}{0.000949}\\
&32 &  0.064 &0.505& 0.544 \\
&03 & 0.104& 0.262& 0.709 \\
\hline
\multirow{3}{*}{$(6,0.3)$}&40 & 0.212& 0.551& 0.558 &\multirow{3}{*}{ 9231.490}&\multirow{3}{*}{\textbf{0.000383 }}\\
&32 & 0.064& 0.505& 0.544 \\
&03 & 0.104& 0.262& 0.709 \\
\hline
\multirow{3}{*}{$(6,0.6)$}&27 & 0.082& 0.254& 0.743 &\multirow{3}{*}{ 8878.365 }&\multirow{3}{*}{0.013094 }\\
&27 &  0.135& 0.474 &0.819 \\
&18 & 0.256& 0.315& 0.466\\
\hline
\multirow{3}{*}{$(6,0.7)$}&18 & 0.053& 0.179& 0.724 &\multirow{3}{*}{  5634.800 }&\multirow{3}{*}{ 0.011617}\\
&18 & 0.113 &0.466& 0.835 \\
&09 & 0.255& 0.315 &0.466 \\
\hline
\end{tabular}}}
\end{center}
\label{b82}
\end{table}

\begin{table}[htb!]
\caption{Optimal Design when $\beta=1$.}
\begin{center}
{\scalebox{0.7}{
\begin{tabular}{ccccccccc}
\hline%
$(\alpha)$&$N_{opt}$&\multicolumn{3}{@{}c@{}}{$\tau_{opt}$}&$C_{opt}$&$Tr[V(\hat{\bm{\theta}})]$\\
\cmidrule(lr){1-1}\cmidrule(lr){2-2}\cmidrule(lr){3-5}\cmidrule(lr){6-6}\cmidrule(lr){7-7}
\multirow{3}{*}{$(-6)$}&40 & 0.212& 0.551& 0.558  &\multirow{3}{*}{9239.490 }&\multirow{3}{*}{\textbf{0.005671 }}\\
&32 &  0.064& 0.505& 0.544  \\
&03 & 0.104& 0.262& 0.709 \\
\hline
\multirow{3}{*}{$(-4)$}&06 & 0.179 &0.564 &0.638  &\multirow{3}{*}{9257.275 }&\multirow{3}{*}{ 0.008650}\\
&19 &  0.137& 0.407 &0.537  \\
&50 & 0.082& 0.434& 0.589 \\
\hline
\multirow{3}{*}{$(-2)$} &40 & 0.212& 0.551& 0.558  &\multirow{3}{*}{ 9239.490}&\multirow{3}{*}{ 0.012310 }\\
&32 &  0.064& 0.505& 0.544  \\
&03 & 0.104& 0.262& 0.709 \\
\hline
\multirow{3}{*}{$(2)$} &40 & 0.212& 0.551& 0.558  &\multirow{3}{*}{9239.490}&\multirow{3}{*}{ 0.009526 }\\
&32 &  0.064& 0.505& 0.544  \\
&03 & 0.104& 0.262& 0.709 \\
\hline
\multirow{3}{*}{$(4)$}&02 & 0.142& 0.449& 0.664  &\multirow{3}{*}{  7811.125}&\multirow{3}{*}{  0.01915}\\
&15 &  0.135& 0.409& 0.546  \\
&46 & 0.113 &0.417 &0.572 \\
\hline
\multirow{3}{*}{$(6)$}&06 & 0.179& 0.564& 0.638  &\multirow{3}{*}{ 9257.275}&\multirow{3}{*}{  0.051036}\\
&19 &   0.137 &0.407& 0.537  \\
&50 & 0.082& 0.434 &0.589  \\
\hline
\end{tabular}}}
\end{center}
\label{b1}
\end{table}

A numerical study is conducted to develop the optimal plan using CPSO by considering a PSALT experiment with stress rates $(3,8,10),$ where NOSD testing units undergo three inspection times at each stress level.  The lifetime of NOSD follows Log-logistic distribution with parameters ($a=1.6, b=1.1, \mu=2.7$).  For constraints setup, the maximum inspection time, experiment budget, and salvage value are taken as $\tau_{max}=1$, $C_r=\$10,000$ and $c_v=\$50,$ respectively, while other cost parameters are adopted from the study of Wu et al. \cite{wu2020optimal} which are $c_a=\$850$, $c_u=\$120$, $c_0=\$55$ and $c_s=\$15.$  

For implementing CPSO, a swarm of $L{=}20$ particles is initialized with initial solution comprising test items \((N_i\,;\,i=1,2,3)\) and inspection times \((\tau_{ij}\,;\,i,j=1,2,3)\), randomly generated within ranges of \((1,75)\) and \((0,1)\), respectively.  Since $N_i$ denotes the number of items put to test, the greatest integer values in \(N_i\) are recorded, and control parameters are set as $w{=}0.3,$ $c_1{=}c_2{=}0.5$ \cite{eltamaly2021novel}.  The experiment terminates, when, for any two consecutive iterations, the difference between $\Phi$ values is less than the threshold value $10^{-8}$ or $t=500$ iterations have reached.  Tables \ref{b0}-\ref{b1} present the optimal designs with different combinations of tuning parameters $(\alpha,\beta,\gamma)$.  These tables consist of optimal sample allocation $N_{opt},$ optimal inspection times $\tau_{opt}$, experimental cost $C_{opt}$ and trace value of the asymptotic covariance matrix of the estimates $Tr[V(\hat{\bm{\theta}})],$ where the minimum $Tr[V(\hat{\bm{\theta}})]$ value is highlighted in bold. 

The analysis across various tables reveals a pattern in the optimal estimation strategies based on the contribution of Bregman exponential divergence (BED) and density power divergence (DPD).  In table \ref{b0} where $\beta{=}0$ indicates the absence of the BED measure, the most efficient strategy relies solely on DPD-based estimation with a tuning parameter $\gamma{=}0.3.$  As the presence of BED increases, the optimal strategies adjust accordingly.  In table \ref{b22} where $\beta{=}0.2$ suggests a lower degree of BED presence, the optimal strategy involves estimates with  $(\alpha{=}-4,\gamma{=}0.3).$  This indicates that even a slight presence of BED can influence the choice of $\alpha.$  Further, in table \ref{b52} where $\beta{=}0.5$ signifies an equal degree of presence of BED and DPD, the optimal strategy shifts towards a positive value of the tuning parameter $\alpha$ and higher value of $\gamma,$ specifically $(\alpha{=}6,\beta{=}0.6).$  However, for a higher degree of BED presence, as seen in table \ref{b82}, where, $\beta{=}0.8$ the optimal plan involves estimation with $(\alpha{=}6,\gamma{=}0.3).$  Interestingly, when DPD is absent $\beta{=}1,$ the most efficient plan reverts to a negative value of $\alpha{=}-6$ based BED estimation.  The key takeaway from these findings is that, under this specific simulation scheme, the estimation with $\gamma=0.3$ mostly provides the best optimal strategy across different degrees of BED presence.

\section{Conclusion}\label{sec7}
This study has developed a robust inferential approach using Exponential-polynomial divergence (EPD) for NOSD testing units under PSALT model.  The asymptotic distribution of the minimum EPD estimate (MEPDE) is derived, with robustness validated through influence function analysis.  A numerical analysis demonstrated the robustness of MEPDE as bias of MEPDE in contamination is less compared to MLE.  Moreover, a concrete score matching (CSM) based approach is proposed for optimizing tuning parameters $(\alpha,\beta,\gamma).$  Comparisons with existing methods indicate that the CSM approach for finding optimal $(\alpha,\beta,\gamma)$ performs efficiently in high contamination.  Further, a data analysis is conducted to validate the theoretical results developed in this study.  Finally, a constraint-based Particle Swarm Optimization (CPSO) method is employed to investigate an optimal experimental design, ensuring efficient sample allocation and inspection time within specified budget and time constraints.  Results show that for the scheme considered in this study, minimum DPD estimate (MDPDE) with tuning parameter $\gamma{=}0.3$ provides the best efficient optimal strategy. 

Simulation experiments and data analysis indicate that no single optimization method consistently identifies optimal tuning parameters in all scenarios.  Thus, a universally effective strategy for tuning parameter selection remains an open challenge.  Additionally, the PSALT model may also be reanalyzed within a competing risk framework.  Ongoing research in these directions will be reported in future studies.

\begin{appendices}
\section{Description of expression $u_{ij}$}
The expressions $u_{ij}$ for $j=1,2,\dots,J_i$ and $u_{i(J_i+1)}=u_{is}$ are described below for the current study.  
 \begin{align*}
        u_{is}&=\frac{\left\{1-p_{is}^{-(b+1)}\right\}}{(b+1)p_{is}^{-(b+1)}}\times
        \begin{bmatrix}
\mu/a\vspace{0.25cm}\\
\frac{1}{b+1}ln\,\{A_2(\bm{\theta};\tau_{iJ_i})\}\vspace{0.25cm}\\
\frac{1}{\mu}ln\,\left\{p_{is}^{-(b+1)}-1\right\}
\end{bmatrix}\\
u_{ij}&=-\big\{p_{ij}(b+1)\big\}^{-1}\times[M]
\end{align*}
\begin{align*}
&[M]=\begin{bmatrix}
\frac{\mu}{a}\Big\{A_3(\bm{\theta};\tau_{i(j-1)})-A_3(\bm{\theta};\tau_{ij})\Big\}\vspace{0.25cm}\\
\frac{1}{b+1}\Big\{A_3(\bm{\theta};\tau_{i(j-1)})\,ln\big(A_2(\bm{\theta};\tau_{i(j-1)})\big)\Big.\\
\quad-\Big.A_3(\bm{\theta};\tau_{ij})\,ln\big(A_2(\bm{\theta};\tau_{ij})\big)\Big\}\vspace{0.25cm}\\
\frac{1}{\mu}\Big\{A_3(\bm{\theta};\tau_{i(j-1)})\,ln\big(A_1(\bm{\theta};\tau_{i(j-1)})\big)\Big.\\
\Big.\quad-A_3(\bm{\theta};\tau_{ij})\,ln\big(A_1(\bm{\theta};\tau_{ij})\big)\Big\}\vspace{0.25cm}
\end{bmatrix}\\
A_1(\bm{\theta};\tau_{ij})&=(a\nu_i^b)^{\mu}\tau_{ij}^{\mu(b+1)}\\
A_2(\bm{\theta};\tau_{ij})&=\frac{(\tau_{ij}\nu_i)^{\mu(b+1)}}{\{1+A_1(\bm{\theta};\tau_{ij})\}^{[1+A_1(\bm{\theta};\tau_{ij})^{-1}]}}\\
A_3(\bm{\theta};\tau_{ij})&=A_1(\bm{\theta};\tau_{ij})\Big\{1+A_1(\bm{\theta};\tau_{ij})\Big\}^{-\frac{b+2}{b+1}}
\end{align*}

\section{Proof of Results}
\subsection{\textbf{Proof of Result \ref{res2}}}
\noindent Define $X_{ui}=(X_{ui1},X_{ui2},\dots,X_{ui(J_i+1)})$ such that $X_{ui}\sim MN(1,\bm{p}_i)$, where, $\bm{p}_i=(p_{i1},p_{i2},\dots,p_{i(J_i+1)}).$ Therefore, $n_{ij}=\sum_{ui=1}^{N_i}X_{uij}.$ Hence, EP divergence ignoring the terms independent of parameters is given as
\begin{align}
    D(\bm{\theta})&=\sum_{i=1}^{k}\frac{1}{N_i}\sum_{ui=1}^{N_i}\left[\sum_{j=1}^{J_i+1}\left\{(1-\beta)p^{\gamma+1}_{ij}+\frac{\beta}{\alpha}e^{\alpha p_{ij}}\left(p_{ij}-\frac{1}{\alpha}\right)\right\}\right.\notag\\
    &\qquad \left.-\sum_{j=1}^{J_i+1}\left\{\frac{\beta}{\alpha}e^{\alpha p_{ij}}+\frac{(\gamma+1)(1-\beta)}{\gamma}p^{\gamma}_{ij}\right\}X_{uij}\right]\notag\\
    &=\sum_{i=1}^{k}\left[\frac{1}{N_i}\sum_{ui=1}^{N_i} V_{ui}(\bm{\theta})\right], \label{veq}
\end{align}
where, 
\begin{align}
   V_{ui}(\bm{\theta})&=\sum_{j=1}^{J_i+1}\left\{(1-\beta)p^{\gamma+1}_{ij}+\frac{\beta}{\alpha}e^{\alpha p_{ij}}\left(p_{ij}-\frac{1}{\alpha}\right)\right\}\\
   &-\sum_{j=1}^{J_i+1}\left\{\frac{\beta}{\alpha}e^{\alpha p_{ij}}+\frac{(\gamma+1)(1-\beta)}{\gamma}p^{\gamma}_{ij}\right\}X_{uij}.\label{vet}
\end{align}
The first and second-order derivatives of equation \eqref{vet} are given below.
\begin{align}
\frac{\partial(V_{ui}(\bm{\theta}))}{\partial\bm{\theta}}&=\sum_{j=1}^{J_i+1}\bigg\{(1-\beta)(\gamma+1)p^{\gamma}_{ij}+\beta p_{ij}e^{\alpha p_{ij}}\bigg\}\frac{\partial(p_{ij})}{\partial\bm{\theta}}\notag\\
&-\sum_{j=1}^{J_i+1}\bigg\{\beta e^{\alpha p_{ij}}+(\gamma+1)(1-\beta)p_{ij}^{\gamma-1}\bigg\}X_{uij}\frac{\partial(p_{ij})}{\partial\bm{\theta}}.\label{fstder}
\end{align}

\begin{align}
\frac{\partial^2 V_{ui}(\bm{\theta})}{\partial\bm{\theta}_l\partial\bm{\theta}_{l^{'}}}&=\sum_{j=1}^{J_i+1}\bigg\{(1-\beta)(\gamma+1)\gamma p_{ij}^{\gamma-1}\bigg.\bigg.+\beta(p_{ij}\alpha e^{\alpha p_{ij}}+e^{\alpha p_{ij}})\bigg\}\frac{\partial(p_{ij})}{\partial\bm{\theta}_l}\frac{\partial(p_{ij})}{\partial\bm{\theta}_{l^{'}}}\notag\\
&+\sum_{j=1}^{J_i+1}\bigg\{(1-\beta)(\gamma+1)p_{ij}^{\gamma}+\beta p_{ij}e^{\alpha p_{ij}}\bigg\}\frac{\partial^2(p_{ij})}{\partial\bm{\theta}_l\partial\bm{\theta}_{l^{'}}}\notag\\
&-\sum_{j=1}^{J_i+1}\bigg\{\beta\alpha e^{\alpha p_{ij}}+(\gamma+1)(1-\beta)(\gamma-1)p_{ij}^{\gamma-2}\bigg\} X_{uij}\frac{\partial(p_{ij})}{\partial\bm{\theta}_l}\frac{\partial(p_{ij})}{\partial\bm{\theta}_{l^{'}}}\notag\\
&-\sum_{j=1}^{J_i+1} \bigg\{\beta e^{\alpha p_{ij}}+(\gamma+1)(1-\beta)p_{ij}^{\gamma-1}\bigg\}X_{uij} \frac{\partial^2(p_{ij})}{\partial\bm{\theta}_l\partial\bm{\theta}_{l^{'}}}.\label{secder}
\end{align}
Let, $Y=\frac{\partial(V_{ui}(\bm{\theta}))}{\partial\bm{\theta}}.$ For multinomial distribution, we have $E(X_{uij})=p_{ij}$, $Var(X_{uij})=p_{ij}(1-p_{ij})$ and $Cov(X_{uij_1},X_{uij_2})=-p_{ij_1}p_{ij_2}.$ Then from equation \eqref{fstder}, $E(Y)=0$ and 
\begin{align}
Var(Y)&=\sum_{j=1}^{J_i+1}\Bigg[\bigg\{\beta e^{\alpha p_{ij}}{+}(\gamma{+}1)(1{-}\beta)p_{ij}^{\gamma{-}1}\bigg\}^2p_{ij}(1{-}p_{ij}) \Bigg.\notag\\
&\Bigg.\left(\frac{\partial(p_{ij})}{\partial\bm{\theta}}\right)^2\Bigg]{-}2\sum_{(j_1,j_2)}^{}\Bigg[\bigg\{\beta e^{\alpha p_{ij_1}}+(\gamma+1) (1-\beta)p_{ij_1}^{\gamma{-}1}\bigg\}\Bigg.\notag\\
&\bigg\{\beta e^{\alpha p_{ij_2}}{+}(\gamma{+}1)(1{-}\beta)p_{ij_2}^{\gamma{-}1}\bigg\}p_{ij_1}p_{ij_2}\frac{\partial(p_{ij_1})}{\partial\bm{\theta}}\frac{\partial(p_{ij_2})}{\partial\bm{\theta}}\Bigg].\label{vare}\\
Cov(Y_l,Y_{l^{'}})&=\sum_{j=1}^{J_i+1}\Bigg[\bigg\{\beta e^{\alpha p_{ij}}+(\gamma+1)(1-\beta)p_{ij}^{\gamma-1}\bigg\}^2 p_{ij}\Bigg.\notag\\
&\Bigg.(1-p_{ij})\frac{\partial(p_{ij})}{\partial\bm{\theta}_l}\frac{\partial(p_{ij})}{\partial\bm{\theta}_{l^{'}}}\Bigg]{-}2\sum_{(j_1,j_2)}^{}\Bigg[\bigg\{\beta e^{\alpha p_{ij_1}}{+}(\gamma{+}1)(1{-}\beta)\Bigg.\bigg.\notag\\
&\quad \Bigg.\bigg.p_{ij_1}^{\gamma-1}\bigg\}\bigg\{\beta e^{\alpha p_{ij_2}}+(\gamma+1)(1-\beta)p_{ij_2}^{\gamma-1}\bigg\}\Bigg. \Bigg.p_{ij_1}p_{ij_2}\frac{\partial(p_{ij_1})}{\partial\bm{\theta}_l}\frac{\partial(p_{ij_2})}{\partial\bm{\theta}_{l^{'}}}\Bigg].\label{cove}
\end{align}
The variance term is $K^{(i)}_{EP}(\bm{\theta})_{jj}=Var(Y)$ and covariance term is $K^{(i)}_{EP}(\bm{\theta})_{jj^{'}}=Cov(Y_j,Y_{j^{'}}).$ Hence,
\begin{align*}
Y=\frac{\partial(V_{ui}(\bm{\theta}))}{\partial\bm{\theta}}&\sim Nor\Big(\bm{0}_3,K^{(i)}_{EP}(\bm{\theta})\Big).\\
\implies \frac{1}{N_i}\sum_{ui=1}^{N_i}\frac{\partial(V_{ui}(\bm{\theta}))}{\partial\bm{\theta}}&\sim Nor\left(\bm{0}_3,\frac{K^{(i)}_{EP}(\bm{\theta})}{N_i}\right).
\end{align*}
Let $H_{\bm{\theta}}=\frac{\partial(D(\bm{\theta}))}{\partial\bm{\theta}}=\sum_{i=1}^{k}\frac{1}{N_i}\sum_{ui=1}^{N_i}\frac{\partial(V_{ui}(\bm{\theta}))}{\partial\bm{\theta}}$ and define $T_{\bm{\theta}}=-\sqrt{N}H_{\bm{\theta}}.$ Then $T_{\bm{\theta}}\sim Nor\big(\bm{0}_3,K_{EP}(\bm{\theta})\big),$ where, $K_{EP}(\bm{\theta})=\sum_{i=1}^{k}K^{(i)}_{EP}(\bm{\theta}).$ Since $\frac{1}{N_i}\sum_{ui=1}^{N_i}X_{uij}\xrightarrow{p} p_{ij}.$ Therefore, from equation \eqref{secder},
\begin{align}
    &\frac{1}{N_i}\sum_{ui=1}^{N_i}\frac{\partial^2 V_{ui}(\bm{\theta})}{\partial\bm{\theta}_l\partial\bm{\theta}_{l^{'}}}\xrightarrow{p}\sum_{j=1}^{J_i+1}\Big[\beta e^{\alpha p_{ij}}+(1-\beta)(\gamma+1)p^{\gamma-1}_{ij}\Big]\frac{\partial(p_{ij})}{\partial\bm{\theta}_l}\frac{\partial(p_{ij})}{\partial\bm{\theta}_{l^{'}}}.
\end{align}
 Hence, 
\begin{align}
&\frac{\partial(H_{\bm{\theta}})}{\partial\bm{\theta}_l}=\frac{\partial^2(D(\bm{\theta}))}{\partial\bm{\theta}_l\partial\bm{\theta}_{l^{'}}}\xrightarrow{p}\sum_{i=1}^{k}\sum_{j=1}^{J_i+1}\Big[\beta e^{\alpha p_{ij}}+(1-\beta)(\gamma+1)p^{\gamma-1}_{ij}\Big]\frac{\partial(p_{ij})}{\partial\bm{\theta}_l}\frac{\partial(p_{ij})}{\partial\bm{\theta}_{l^{'}}}.\label{all}
\end{align}
If $\theta^0$ represents true values of parameters, then by Taylor series expansion ignoring higher order terms,
\begin{align*}
H_{\bm{\theta}}&=H_{\bm{\theta}^0}+\sum_{l=1}^{3}\left.\frac{\partial(H_{\bm{\theta}})}{\partial\bm{\theta}_l}\right\vert_{\bm{\theta}=\bm{\theta}^0}(\hat{\bm{\theta}}_l-\bm{\theta}^0_l)+\frac{1}{2}\sum_{l=1}^{3}\sum_{l^{'}=1}^{3}\left.\frac{\partial^2(H_{\bm{\theta}})}{\partial\bm{\theta}_l\partial\bm{\theta}_{l^{'}}}\right\vert_{\bm{\theta}=\bm{\theta}^0}(\hat{\bm{\theta}}_l-\bm{\theta}^0_l)(\hat{\bm{\theta}}_{l^{'}}-\bm{\theta}^0_{l^{'}}).
\end{align*}
Since $H_{\hat{\bm{\theta}}}=0$, therefore,
\begin{equation}
    -\sqrt{N}H_{\bm{\theta}^0}=\sqrt{N}\sum_{l=1}^{3}(\hat{\bm{\theta}}_l-\bm{\theta}_l^0) A_{(l,l^{'})},\label{tal}
\end{equation}
\begin{equation*}
 \text{where,} A_{(l,l^{'})}{=}\left.\frac{\partial(H_{\bm{\theta}})}{\partial\bm{\theta}_l}\right\vert_{\bm{\theta}{=}\bm{\theta}^0} {+}\frac{1}{2}\sum_{l=1}^{3}\sum_{l^{'}=1}^{3}\left.\frac{\partial^2(H_{\bm{\theta}})}{\partial\bm{\theta}_l\partial\bm{\theta}_{l^{'}}}\right\vert_{\bm{\theta}{=}\bm{\theta}^0}(\hat{\bm{\theta}}_{l^{'}}{-}\bm{\theta}^0_{l^{'}})  
\end{equation*}
 and $A_{(l,l^{'})}\xrightarrow{p}\eqref{all}$ implies $A_{\bm{\theta}}\xrightarrow{p}J_{EP}(\bm{\theta}^0),$ where,
\begin{align}
   J_{EP}(\bm{\theta}^0)&=\left[\left(\sum_{i=1}^{k}\sum_{j=1}^{J_i+1}\Big\{\beta e^{\alpha p_{ij}}+(1-\beta)(\gamma+1)p_{ij}^{\gamma-1}\Big\}\right.\right. \left.\left.\frac{\partial(p_{ij})}{\partial\bm{\theta}_l}\frac{\partial(p_{ij})}{\partial\bm{\theta}_{l^{'}}}\right)_{l,l^{'}}\right]. \label{jeq}
\end{align}
Let, $Z_l=\sqrt(N)(\hat{\bm{\theta}_l}-\bm{\theta}_l^0).$ Then by equation \eqref{tal}, $-\sqrt{N}H_{\bm{\theta}^0}=Z_l A_{(l,l^{'})}\implies T_{\bm{\theta}}=Z_{\bm{\theta}}A_{\bm{\theta}}\implies Z_{\bm{\theta}}=A_{\bm{\theta}}^{-1}T_{\bm{\theta}}.$ Therefore,
\begin{equation}
    \sqrt{N}(\hat{\bm{\theta}}{-}\bm{\theta}^0){=}A_{\bm{\theta}}^{-1}T_{\bm{\theta}}{\sim} Nor\Big(\bm{0}_{3}, J^{-1}_{\scriptscriptstyle EP}({\bm{\theta}_0}) K_{\scriptscriptstyle EP}({\bm{\theta}_0})J^{-1}_{\scriptscriptstyle EP}({\bm{\theta}_0}) \Big).
\end{equation}
\subsection{\textbf{Proof of Result \ref{res3}}}
The influence function (IF) is obtained following the procedure adopted by Ghosh and Basu \cite{ghosh2013robust} and Castilla and  Chocano \cite{castilla2022new}.  Let $G_i$ be the true distribution with probability mass function (pmf) $g_i$ for ith group with $i=1,\dots,k$ and denote $\bm{G}=(G_1,\dots,G_i,\dots,G_k)^T$ . 
If $T_{EP}(G)$ denotes the statistical functional of $\hat{\theta}_{EP},$ then $T_{EP}(G)$ will satisfy
\begin{equation}
\sum_{i=1}^{k}\bm{\pi}^T_{i}(\bm{\theta})\Big\{g_{i}-\bm{p}_{i}(\bm{\theta})\Big\}=0,\label{functional}
\end{equation}
where, the vectors, $\bm{\pi}_i=(\pi_{i1},\dots,\pi_{iJ_i+1})^T$ with $\pi_{ij}(\bm{\theta})=\Big\{(1-\beta)(\gamma+1)p^{\gamma-1}_{ij}(\bm{\theta})+\beta\,e^{\alpha p_{ij}(\bm{\theta})}\Big\}\frac{\partial(p_{ij}(\bm{\theta}))}{\partial \theta}$ and $\bm{p}_i=(p_{i1},\dots,p_{iJ_i+1}).$  In the $i$th group $i=1,\dots,k$ for any $l$th unit $l=1,\dots,N_i$, we observe $J_i+1$ dimensional vector,
$$y_{il}{=}(y_{i1l},{\dots},y_{iJ_i+1})^T\;;\;i{=}1,2,{\dots},k\;;\;j=1,2,{\dots},J_i{+}1,$$
where,
$$y_{ijl}{=} 
\begin{cases}
1 \; \;;\;  \text{ if $l$th unit of $i$th group fails in $(\tau_{i(j-1)},\tau_{ij}]$ }\\
0 \;\;;\quad \text{otherwise}
\end{cases}.
$$\\ 
Further, consider the contaminated probability vector 
$$g_{i,\epsilon}=(1-\epsilon)\bm{p}_i(\bm{\theta}^0)+\epsilon \delta_{\bm{t_i}},$$
where, $\epsilon$ is the contamination proportion and $\delta_{\bm{t_i}}$ is the degenerate probability at the outlier point $\bm{t_i}=(t_{i1},\dots,t_{iJ_i+1})^T\in\{0,1\}^{J_i+1}$ with $\sum_{j=1}^{J_i+1}t_{ij}=1.$  Let us assume that the contamination is only in $i_0$ group.  Therefore,
$$g_i = 
\begin{cases}
\bm{p}_i(\bm{\theta}^0) \quad \text{if} \  i\neq i_0\\
g_{i_0,\epsilon} \qquad \text{if} \ i=i_0\\
\end{cases}.
$$\\
Then replacing $\bm{\theta}$ by $\bm{\theta}^{i_0}_{\epsilon}$ and $g_{i_0}$ by $g_{i_0,\epsilon}$, the equation \eqref{functional} takes the form
\begin{align}
&\sum_{\substack{i=1 \\ i \neq i_0}}^{k}\bm{\pi}^T_i(\bm{\theta}^{i_0}_{\epsilon})\bm{p}_i(\bm{\theta}^0)+\bm{\pi}^T_{i_0}(\bm{\theta}^{i_0}_{\epsilon})\Big[(1-\epsilon)\bm{p}_{i_0}(\bm{\theta}^0)+\epsilon \delta_{\bm{t_{i_0}}}\Big]\notag\\
&\qquad\qquad-\sum_{i=1}^{k}\bm{\pi}^T_i(\bm{\theta}^{i_0}_{\epsilon})\bm{p}_i(\bm{\theta}^{i_0}_{\epsilon})=0.\label{confun}
\end{align}
Differentiating equation \eqref{confun} with respect to $\epsilon$ and taking $\epsilon\to 0^{+},$ we get the expression for IF,
\begin{equation}
IF(\bm{t_{i_0}},T_{EP},G){=}J^{-1}_{EP}(\bm{\theta}^0) \sum_{j=1}^{J_i+1}\bm{\pi}_{i_0j}(\bm{\theta}^0)\left\{\delta_{\bm{t_{i_0}}}{-}p_{i_0j}(\bm{\theta}^0)\right\}.
\end{equation}
Considering contamination across all groups, the IF can be obtained as,
\begin{align}
IF(\bm{t_1},\dots,\bm{t_k},T_{EP},G)&=J^{-1}_{EP}(\bm{\theta}^0) \sum_{i=1}^{k}\sum_{j=1}^{J_i+1}\pi_{ij}(\bm{\theta}^0)\left\{\delta_{\bm{t_i}}-p_{ij}(\bm{\theta}^0)\right\}.
\end{align}
\end{appendices}
{\scriptsize
\bibliographystyle{elsarticle-num}
 \bibliography{cas-refs}}
\end{document}